\documentclass[aps,prd,twocolumn,showpacs]{revtex4}
\usepackage[hyperindex,breaklinks]{hyperref}
\usepackage{graphicx}
\def\figvolexp{\centerline{\scalebox{0.3}{\includegraphics{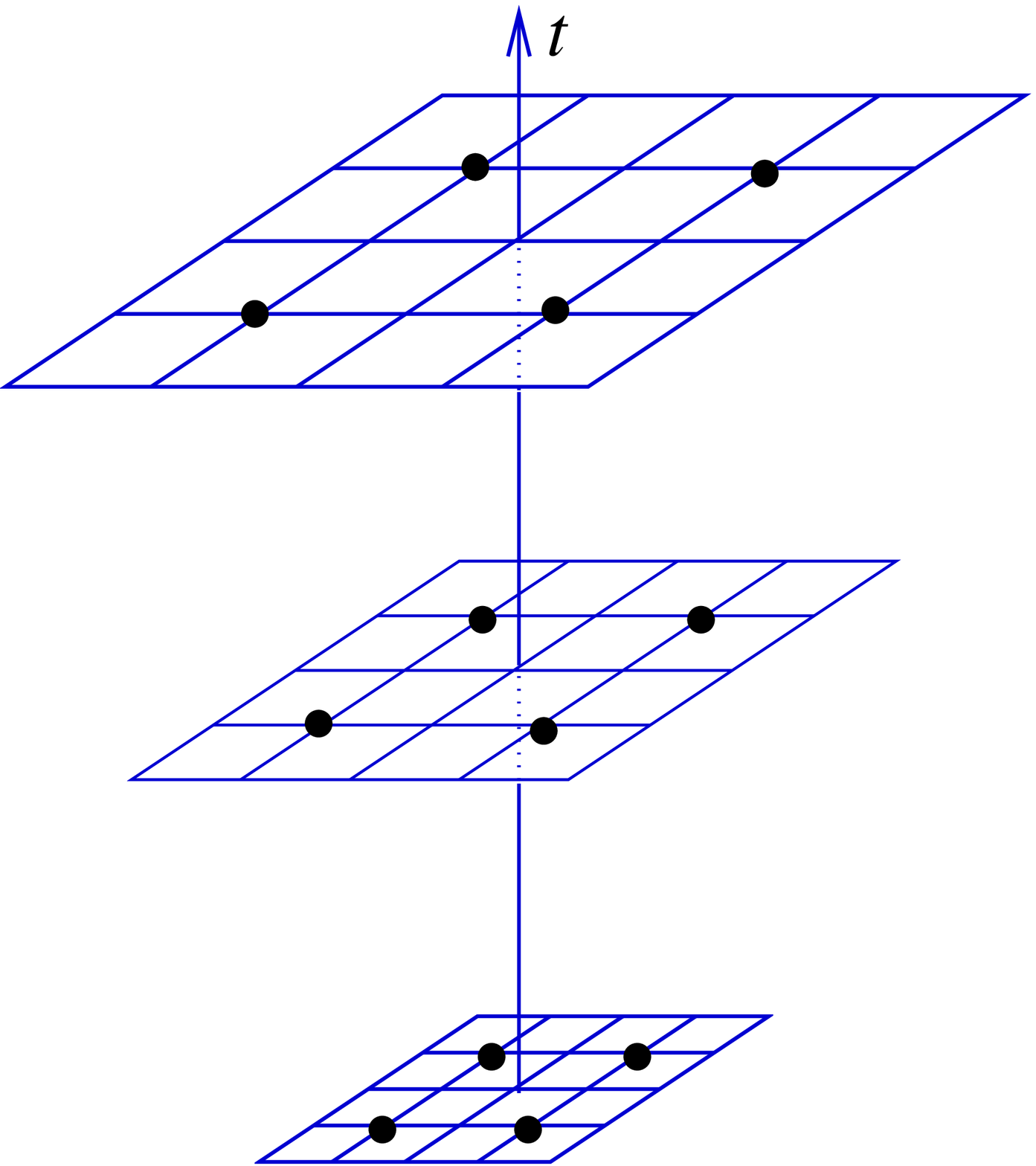}}}}
\def\figlattice{\centerline{\scalebox{0.3}{\includegraphics{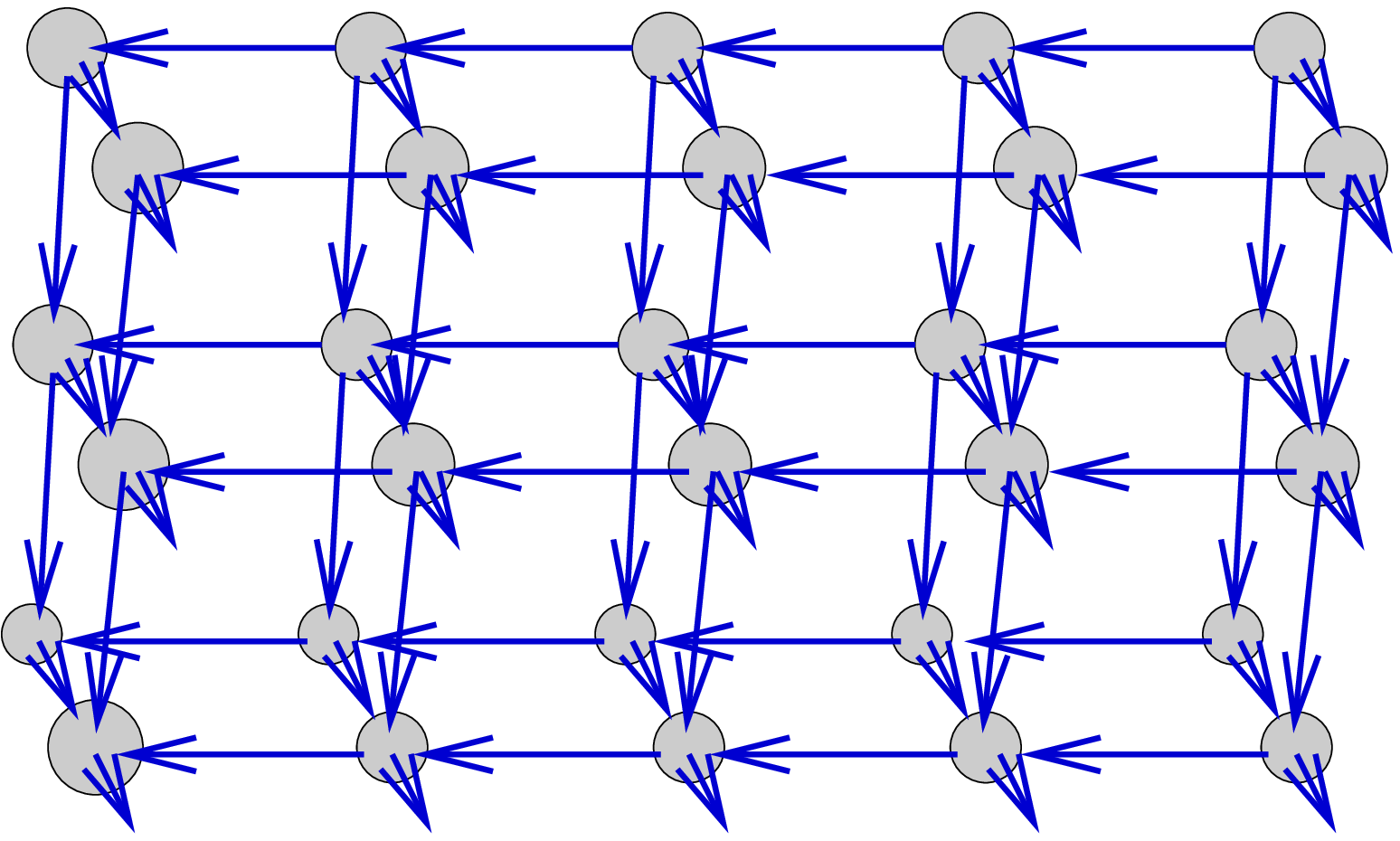}}}}
\def\figequivM{\centerline{\scalebox{0.35}{\includegraphics{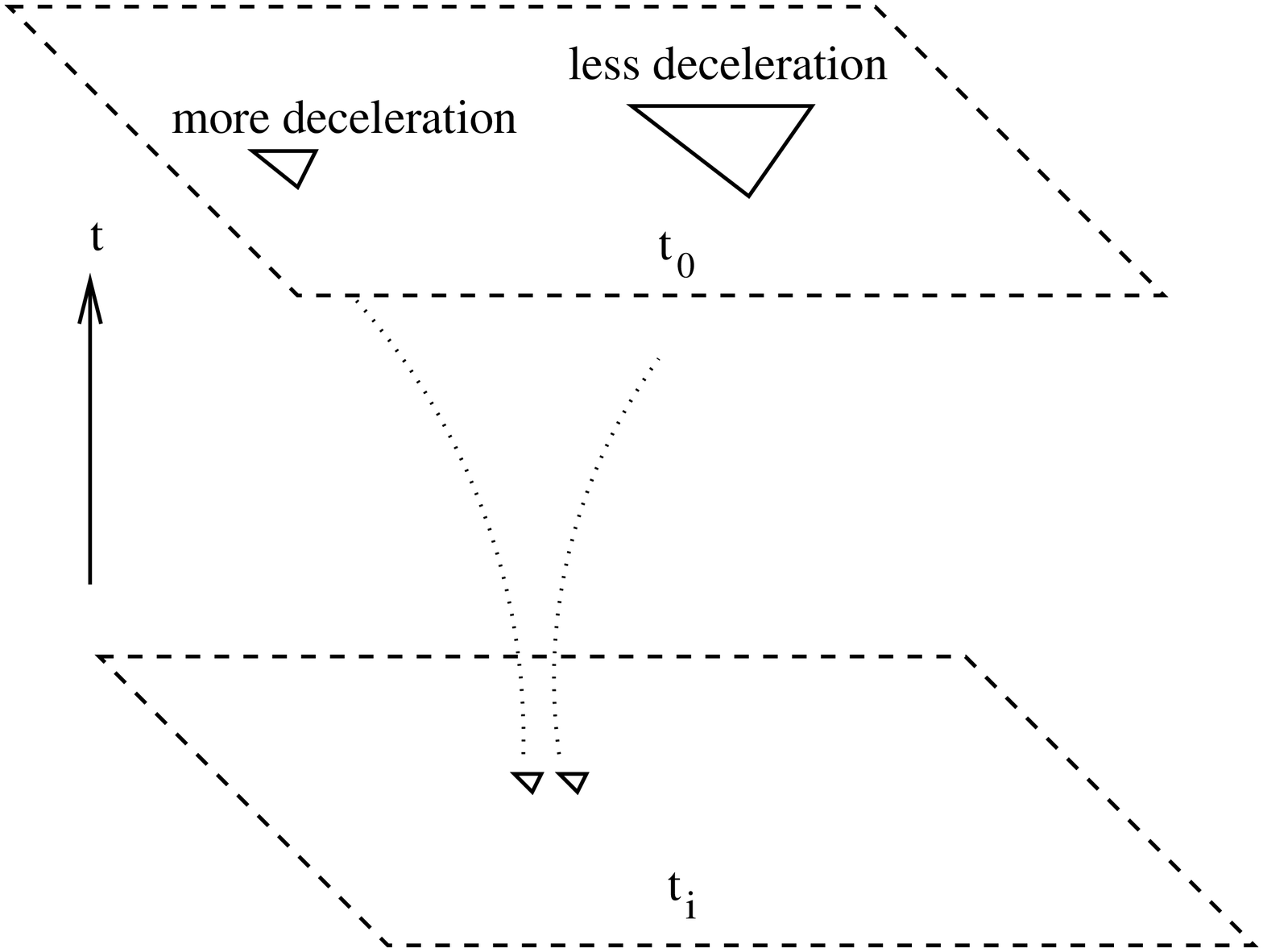}}}
\vskip-20pt\leftline{\bf(a)}\vskip25pt}
\def\figequivGR{\centerline{\scalebox{0.35}{\includegraphics{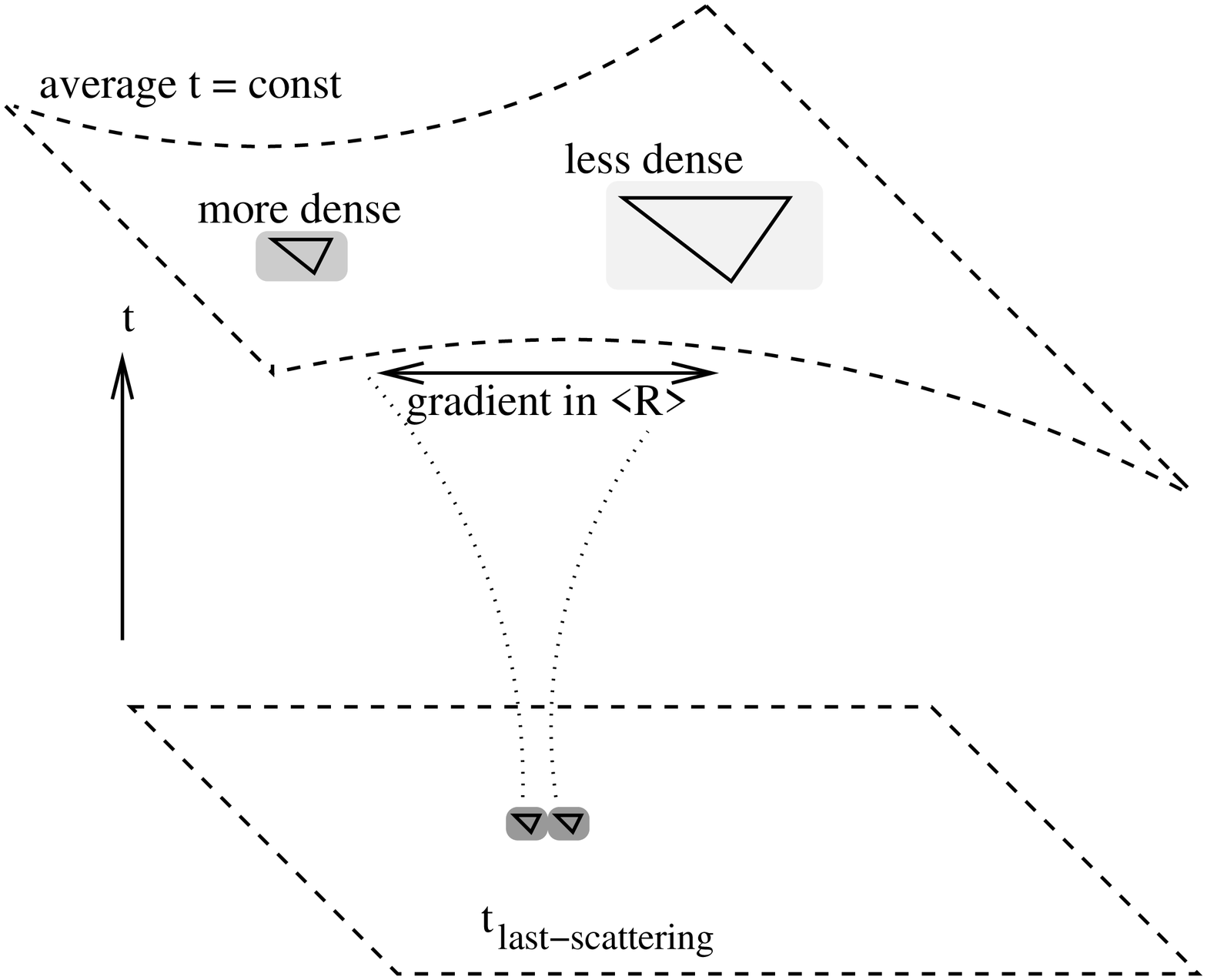}}}
\vskip-20pt\leftline{\bf(b)}\vskip25pt}
\def\figaccela{\centerline{\scalebox{0.5}{\includegraphics{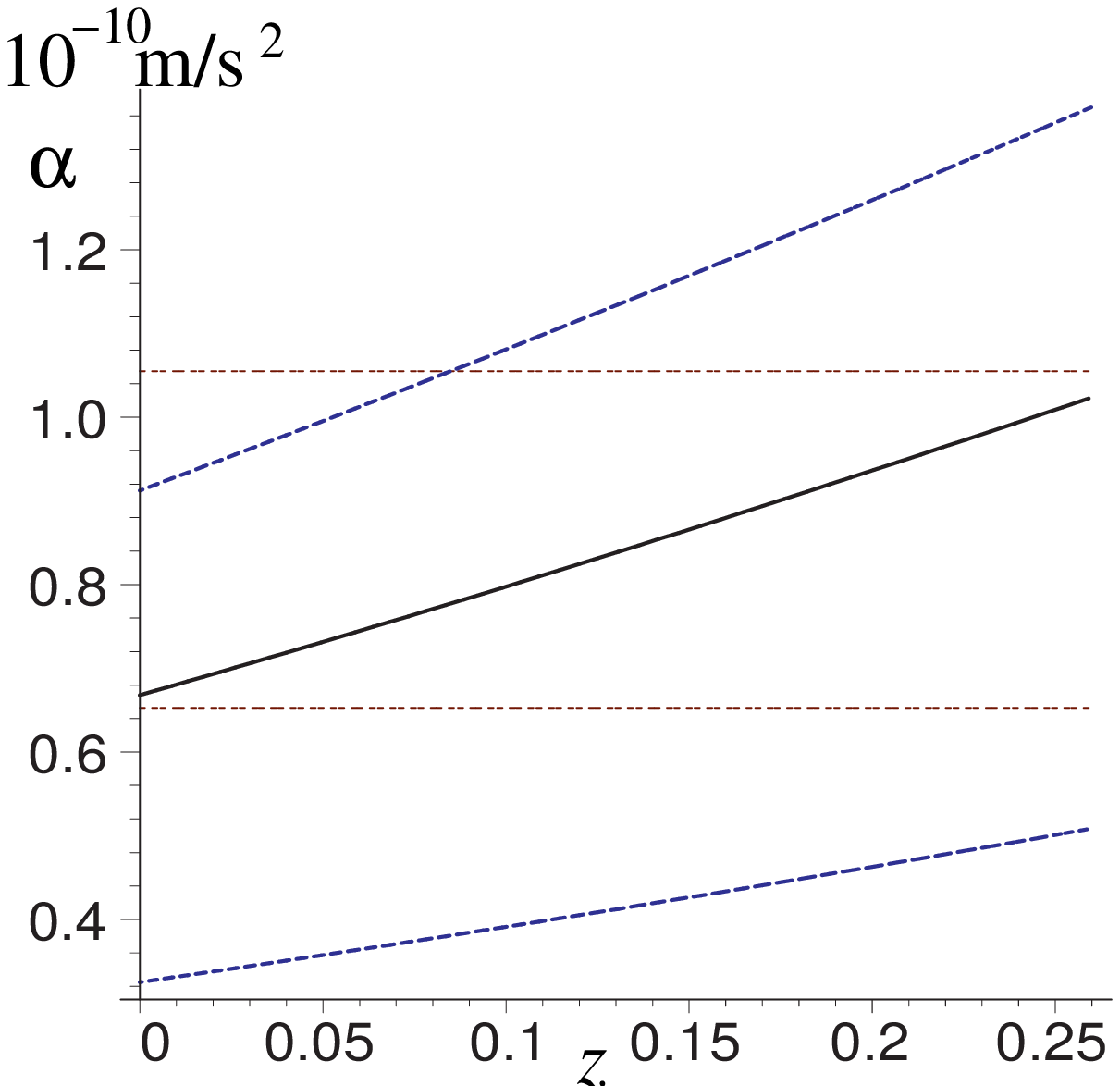}}}
\vskip-20pt\leftline{\bf(a)}\vskip10pt}
\def\figaccelb{\centerline{\scalebox{0.5}{\includegraphics{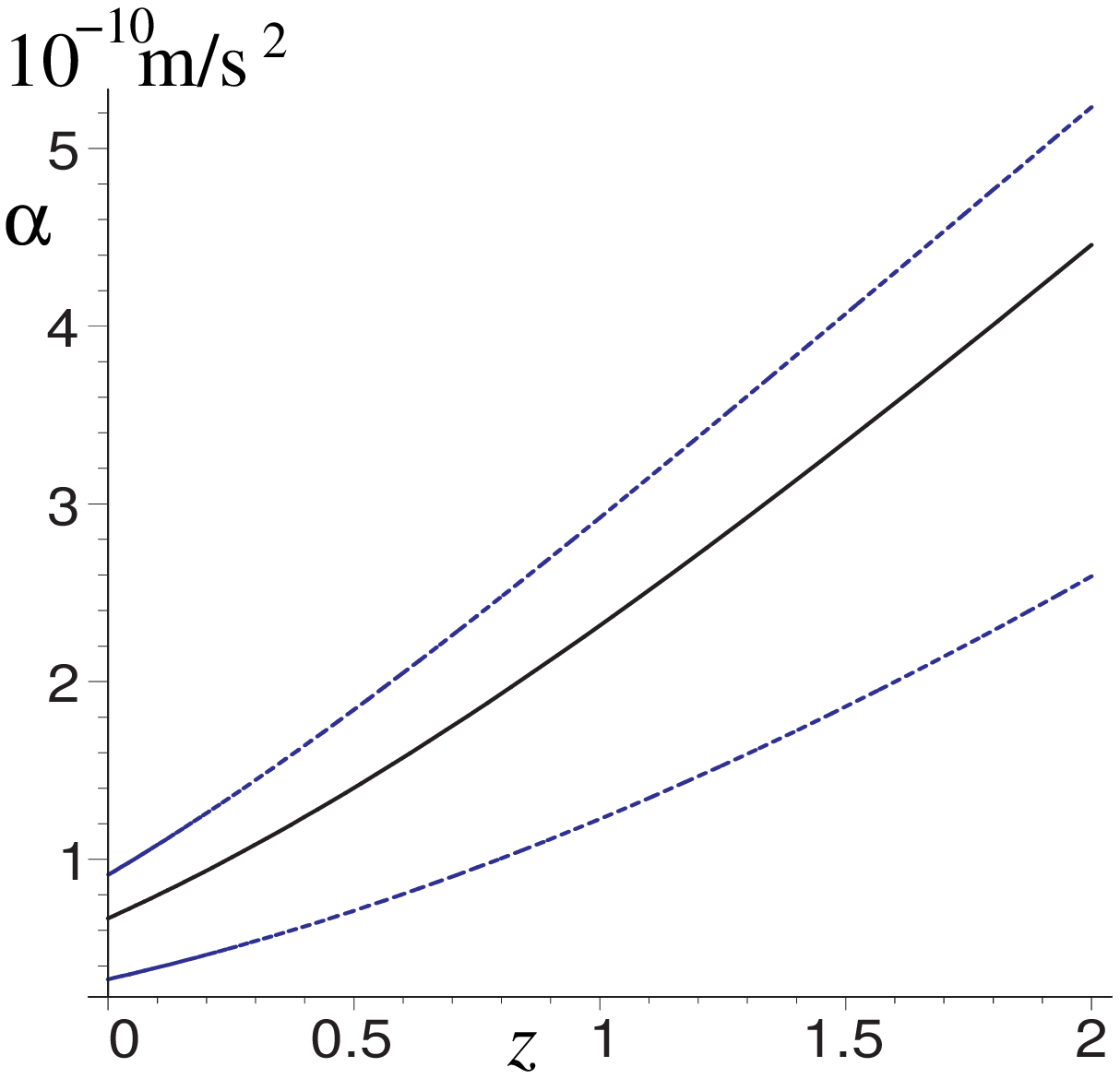}}}
\vskip-20pt\leftline{\bf(b)}\vskip5pt}
\def\figaccelHc{\centerline{\scalebox{0.5}{\includegraphics{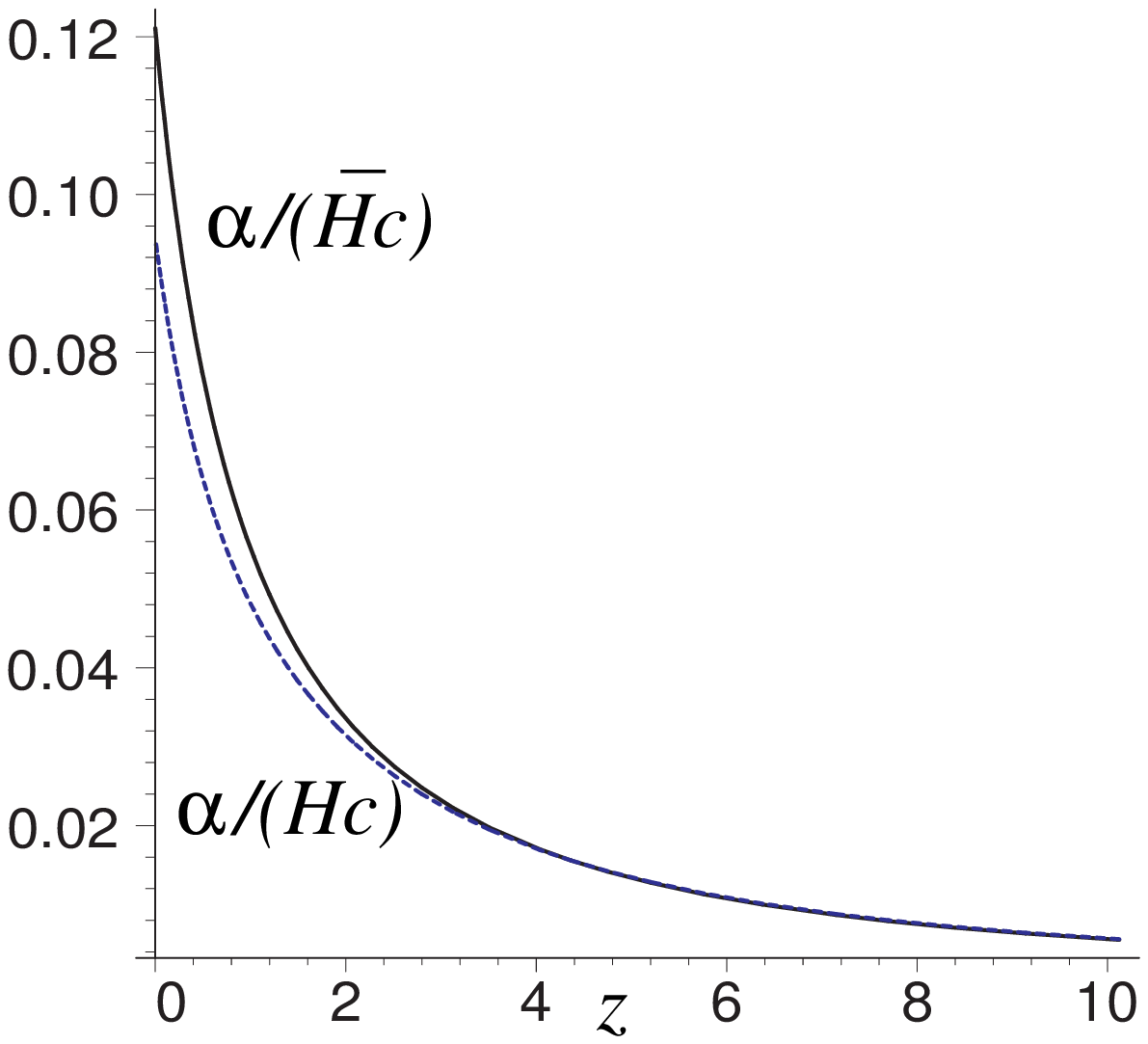}}}}

\begin{document}
\def\PRL#1{Phys.\ Rev.\ Lett.\ {\bf#1}} \def\PR#1{Phys.\ Rev.\ {\bf#1}}
\def\ApJ#1{Astrophys.\ J.\ {\bf#1}} \def\PL#1{Phys.\ Lett.\ {\bf#1}}
\def\MNRAS#1{Mon.\ Not.\ R.\ Astr.\ Soc.\ {\bf#1}} 
\def\CQG#1{Class.\ Quantum Grav.\ {\bf#1}}
\def\GRG#1{Gen.\ Relativ.\ Grav.\ {\bf#1}}
\def\AaA#1{Astron.\ Astrophys.\ {\bf#1}}
\def\beq{\begin{equation}} \def\eeq{\end{equation}}
\def\bea{\begin{eqnarray}} \def\eea{\end{eqnarray}}
\def\Z#1{_{\lower2pt\hbox{$\scriptstyle#1$}}} \def\w#1{\,\hbox{#1}}
\def\X#1{_{\lower2pt\hbox{$\scriptscriptstyle#1$}}}
\font\sevenrm=cmr7 \def\ns#1{_{\hbox{\sevenrm #1}}} \def\dOM{\dd\Omega^2}
\def\Ns#1{\Z{\hbox{\sevenrm #1}}} \def\ave#1{\langle{#1}\rangle}
\def\lsim{\mathop{\hbox{${\lower3.8pt\hbox{$<$}}\atop{\raise0.2pt\hbox{$\sim$}}
$}}} \def\kmsMpc{\w{km}\;\w{sec}^{-1}\w{Mpc}^{-1}} \def\bn{\bar n}
\def\dd{{\rm d}} \def\ds{\dd s} \def\etal{{\em et al}.}
\def\al{\alpha}\def\be{\beta}\def\ga{\gamma}\def\de{\delta}\def\ep{\epsilon}
\def\et{\eta}\def\th{\theta}\def\ph{\phi}\def\rh{\rho}\def\si{\sigma}
\def\ta{\tau} \def\OM{\widetilde\Omega} \def\deti{\widetilde\de}
\def\ati{\widetilde a}\def\rti{\widetilde r}\def\eti{\widetilde\et}
\def\frn#1#2{{\textstyle{#1\over#2}}} \def\Deriv#1#2#3{{#1#3\over#1#2}}
\def\Der#1#2{{#1\hphantom{#2}\over#1#2}} \def\pt{\partial} \def\ab{{\bar a}}
\def\goesas{\mathop{\sim}\limits} \def\tv{\ta\ns{v}} \def\tw{\ta\ns{w}}
\def\gb{\bar\ga}\def\omi{\OM_i} \def\I{{\hbox{$\scriptscriptstyle I$}}}
\def\av{{a\ns{v}\hskip-2pt}} \def\aw{{a\ns{w}\hskip-2.4pt}}\def\Vav{{\cal V}}
\def\DD{{\cal D}}\def\gd{{{}^3\!g}}\def\half{\frn12}\def\Rav{\ave{\cal R}}
\def\QQ{{\cal Q}}\def\dsp{\displaystyle}\def\bx{{\mathbf x}}\def\bH{\bar H}
\def\bk{{\mathbf k}}\def\bU{{\mathbf U}}\def\cd{\!\cdot\!}
\font\bm=cmmib10 \def\alB{\hbox{\bm\char11}} \def\xiB{\hbox{\bm\char24}}
\def\fv{{f\ns v}}\def\Hb{\bH\Z{\!0}}\def\fvn{f\ns{v0}} \def\OmMn{\Omega\Z{M0}}
\def\la{\lambda}\def\Omb{\bar\Omega\X M}
\title{Cosmological equivalence principle and the weak--field limit}
\author{David L. Wiltshire}
\email{David.Wiltshire@canterbury.ac.nz}
\affiliation{Department of Physics \& Astronomy, University of Canterbury,
Private Bag 4800, Christchurch 8140, New Zealand\footnote{Permanent address};}
\affiliation{International Center for Relativistic Astrophysics Network
(ICRANet), P.le della Repubblica 10, Pescara 65121, Italy}
\begin{abstract}
The strong equivalence principle is extended in application to averaged
dynamical fields in cosmology to include the role of the average density in
the determination of inertial frames. The resulting cosmological equivalence
principle is applied to the problem of synchronisation of clocks in the
observed universe.
Once density perturbations grow to give density contrasts of order one on
scales of tens of megaparsecs, the integrated deceleration of the local
background regions of voids relative to galaxies must be accounted for in
the relative synchronisation of clocks of ideal observers who measure an
isotropic cosmic microwave background. The relative deceleration of the
background can be expected to represent a scale in which weak--field
Newtonian dynamics should be modified to account for dynamical gradients in
the Ricci scalar curvature of space. This acceleration scale is estimated
using the best--fit nonlinear bubble model of the universe with
backreaction. At redshifts $z\lsim0.25$ the scale is found to coincide with
the empirical acceleration scale of modified Newtonian dynamics. At larger
redshifts the scale varies in a manner which is likely to be important for
understanding dynamics of galaxy clusters, and structure formation.
Although the relative deceleration, typically of order $10^{-10}$ms$^{-2}$,
is small, when integrated over the lifetime of the universe it amounts to
an accumulated relative difference of 38\% in the rate of average clocks in
galaxies as compared to volume--average clocks in the emptiness of voids.
A number of foundational aspects of the cosmological equivalence principle
are also discussed, including its relation to Mach's principle, the Weyl
curvature hypothesis and the initial conditions of the universe.
\end{abstract}
\pacs{04.20.Cv, 95.36.+x, 98.80.Jk}
\maketitle
\section{Introduction}

The strong equivalence principle (SEP) stands at the conceptual core of
general relativity, as a physical principle which limits the choice of our
physical theory of gravitation among all possible metric
theories of gravitation one can construct. In this paper I will argue that
the ramifications of this principle have not been fully explored, and
that its physical interpretation requires further clarification to deal
with the dynamical properties of spacetime inherent in Einstein's theory
when the nonequilibrium situation is considered. In
particular, the problem of how to synchronise clocks in the absence of a
spacetime background with specific symmetries does not have a general
solution in general relativity. In this paper I will show that
at least for universes which began with a great deal of symmetry, as ours
did, the reasoning of the equivalence principle can be extended:
the average regional density provides a clock in expanding regions.
This has particular consequences for cosmological models and the definition
of gravitational energy. It underlies the author's proposal that dark
energy is a misidentification of cosmological gravitational energy
gradients in an inhomogeneous void--dominated universe
\cite{clocks,sol,essay}. Broader foundational consequences may also follow.

To set the scene, it pays to recall that historically the equivalence
principle\cite{eep}, and indeed general relativity\cite{GR}, was formulated
at a time before the dynamical properties of spacetime were understood. The
conceptual route that Einstein took began with the {\em weak equivalence
principle} or the {\em principle of uniqueness of free fall}, known
since the experiments of Galileo, that all bodies (subject to no forces
other than gravity) will follow the same paths given the same initial
positions and velocities. Realising that this observational statement
embodies a feature of universality of the gravitational interaction,
Einstein created a theory in which gravity is a property of spacetime
itself.

Einstein's identification of what the true gravitation field should be began
101 years ago with first identifying what it is not, based on thought
experiments with elevators, concerning what motions of particles cannot be
distinguished operationally. His 1907 principle of equivalence \cite{eep}
may be translated as follows: {\em All motions in an external static
homogeneous gravitational field are identical to those in no gravitational
field if referred to a uniformly accelerated coordinate system}. A uniformly
accelerated reference frame may be found operationally in empty
Minkowski spacetime by firing rockets; if matter is the source of the
true gravitational field then such choices of frame cannot represent
gravity.

More generally, since special relativity with nongravitational forces
appears to always be valid in small regions, one should always be able
to get rid of gravity near a point. This is embodied in the SEP: {\em At
any event, always and everywhere, it is possible to choose a local inertial
frame such that in a sufficiently small spacetime neighbourhood all
nongravitational laws of nature take on their familiar forms appropriate to
the absence of gravity, namely the laws of special relativity}. This means
that gravity is made to be universal, as it is contained in spacetime
structure. The true gravitational field strength is encoded in the Riemann
curvature tensor, and is determined regionally by the tidal effects of
geodesic deviation.

One of the most profound and difficult consequences of the SEP is that
gravitational energy and momentum cannot be described by a local density,
and so are not local quantities.
General relativity overcomes the nonlocality problem of Newtonian
gravity: there is no action at a distance. General relativity is an
entirely local theory in the sense of {\em propagation} of the gravitational
interaction, which is causal. However, the background on which the
interaction propagates may contain its own energy and momentum, when
integrated over sufficiently large regions, and this has
to be understood in the calibration of local rods and clocks at
widely separated events.

Whereas the calibration of rods and clocks is mathematically determined by
invariants of the {\em local} metric, and the spacetime connection which
relates local invariants at widely separated events, in practice we cannot
analytically solve Einstein's equations for the most general distribution
of matter to unambiguously determine the metric and its connection. A
slicing of spacetime into hypersurfaces, and a threading of these hypersurfaces
by timelike worldlines of observers or by null geodesics, is inevitable
for any operational description of spacetime in terms of rods and
clocks. Such splittings of space and time, together with additional
symmetry assumptions, are also necessary for analytically solving
Einstein's equations in particular cases, or more generally for numerical
modelling.

The definition of quasilocal gravitational energy and momentum \cite{quasi_rev}
then turns out to depend on the choice of slicing, associated surfaces of
integration, and the identification of observers that thread the slices. These
procedures are inherently noncovariant and nonunique, and many questions
of naturalness of any particular definition inevitably arise. (See Ref.\
\cite{NSV} for a recent discussion.) We have the dilemma that a spacetime
split inevitably breaks any given particle motion into a motion of the
background and a motion with respect to the background; and this may
involve a degree of arbitrariness. The
viewpoint that will be adopted here is that since quasilocal gravitational
energy gradients have their origin in the equivalence principle, the primary
criterion for making canonical identifications of physically relevant
classes of observer frames is that the equivalence principle itself must
be properly formulated, and respected, when making macroscopic cosmological
averages.

If the SEP is applied to macroscopic objects then strictly speaking we
can only apply it to systems in which the gravitational interaction
is ignored. Yet we implicitly apply the SEP to scales
at least as large as galaxies which are treated as particles of dust in
an expanding fluid -- with an expansion rate given by a
Friedmann--Lema\^{\i}tre--Robertson--Walker (FLRW) model --
subject to additional Newtonian gravitational interactions. The
Newtonian approximation, which is made in present--day numerical simulations
of structure formation must, by the rules of general relativity, presuppose
that a sufficiently large static Minkowski frame can be found.

I shall argue in this paper that to correctly derive a Newtonian limit,
without prior assumptions about the cosmological background geometry, we
must first correctly apply the SEP. If galaxies are to be treated as
particles of dust we must address the following question: given that
the background is not static, what is the largest scale on which the SEP
can be applied?
An attempt to answer this question, which is unavoidable if we are
to consistently apply the principles of general relativity to cosmology,
means that we have to deal with the relationship of inertial frames
to averages of matter fields and motions. In other words we must
address Mach's principle, which may be stated as \cite{Bondi,BKL} follows:
{\em``Local inertial frames are determined through the distributions of
energy and momentum in the universe by some weighted average of the apparent
motions''}. The leading questions in this statement are what is to
be understood by ``local'', and what is the ``suitable weighted average''?

The problem with any process of averaging in general relativity is
that by coarse graining we can lose information about the calibration
of local rods and clocks within the coarse grained cells relative to
average quantities. Rods and clocks are related to invariants of the
local metric, can vary greatly within an averaging cell, and will
not in general coincide with some average calibration of rods and
clocks once averaging cells become sufficiently large in the nonlinear
regime of general relativity. Galaxies, for example, contain supermassive
black holes, in whose local neighbourhood the determination of proper
lengths and times for typical observers differs extremely from typical
observers in the outskirts of a galaxy.

It is commonly believed that as long as we are ``in the weak--field limit''
we do not have to worry about complications such as the extreme
ones posed by black holes. However, the weak--field limit is always
taken about a {\em background}, and once inhomogeneities develop in the
universe there are no exact symmetries to describe the background. One set of
uniformly calibrated rods and clocks is no longer sufficient to describe
the background itself. Given an initially homogeneous
isotropic universe with small scale--invariant density perturbations,
once one is within the nonlinear regime of structure
formation, homogeneity and isotropy are only defined in a statistically
average sense.

Within a cell of statistical homogeneity -- which can be taken to be
\cite{clocks} of the same order as the baryon acoustic oscillation
(BAO) scale, $100h^{-1}$Mpc, $h$ being the dimensionless
parameter related to the Hubble constant by $H\Z0=100h\kmsMpc$ -- there are
density contrasts of order unity over scales of tens of megaparsecs.
Since the universe is inhomogeneous over these scales {\em not every observer
is the same average observer}. Different classes of canonical observers
are required to interpret cosmological parameters \cite{clocks}. In
particular, we and the objects we see are typically in galaxies in locally
nonexpanding regions which formed from density perturbations which
were greater than the critical density. In such regions local spatial
curvature can differ markedly from the volume--average location
in voids, where space is freely expanding. Small differences of spatial
curvature and rates of clocks can accumulate to give
large differences over the lifetime of the universe
\cite{clocks}. Dynamical gradients in the curvature of space are a
physical reality which must be properly understood.

In this paper I will estimate the scale of the small relative deceleration
of the background in regions of different density, by proposing an
extension of the SEP to cosmology. I will demand an equivalence between
the particular example of geodesic flow of a congruence of particles
``at rest'' in a dynamically expanding universe, whose average volume
expansion is governed by an average over all masses and motions
of the particles, and the equivalent ``volume--expanding'' motion of
point particles in a Minkowski space. In other words, the local Ricci
scalar curvature due to the volume average of pressureless dust can always be
``renormalised away'' on sufficiently small scales, but in a way which may
lead to relative recalibrations of local rods and clocks between different
spacetime regions.

Equivalently, for volume expansion we cannot {\em locally}
distinguish between particles at rest in a dynamic spacetime, and particles
moving in a static spacetime: the two situations are equivalent in a sense
which deserves the designation {\em cosmological principle of equivalence}.
This might be viewed as a further Machian style refinement of the notion of
inertia. Although we measure geodesic deviation, in terms of the scalar
curvature part of the Riemann tensor and volume expansion, we are unable to
distinguish whether the geodesics are deviating because of local
accelerations of particles in a static space, or whether the particles
are ``at rest'' in an expanding space which is decelerating due to
the gravitational attraction of the average density of matter.

Historically, it might be said
that although Einstein was conceptually guided by Mach's principle,
he never quite succeeded in fully implementing it in general relativity,
because when he first formulated the theory he did not fully appreciate
the dynamical nature of spacetime. His first attempt to study cosmology
indicated that for any usual source of matter the theory was not stable,
but intrinsically dynamical \cite{esu}.
Famously, he invoked the cosmological constant -- his {\em``gr\"o{\ss}te
Eselei''} -- to avoid the issue. This paper attempts to lay the conceptual
groundwork for an alternative first principles route to the physical
interpretation of cosmological general relativity, taking account of
observational evidence that is immeasurably better now than it was in 1917.

The plan of the paper is as follows. Preliminary definitions, motivations,
and a statement of the cosmological equivalence principle
are presented in Sec.\ \ref{aep}. In Sec.\ \ref{gedanken} the thought
experiments introduced in Sec.\ \ref{aep} are generalised to the case
of regions of different density, and an estimate of clock rate variance
is given based on presently observed density contrasts. The definition
of the cosmic rest frame, and its relation to other frames used in
cosmological averaging, is clarified in Sec.\ \ref{frame}. In Sec.\ \ref{alpha}
a numerical estimate of the time--varying relative deceleration of voids
relative to walls, where galaxies are located, is computed over the lifetime
of the universe, and its cosmological implications discussed.
The role of initial conditions and a possible conceptual relationship
to Penrose's Weyl curvature hypothesis are discussed in Sec.\ \ref{Weyl}.
The paper concludes with a summary discussion, including some further
speculations, in Sec.\ \ref{conclude}.

\section{Averaging and the equivalence principle\label{aep}}

The SEP addresses physics on relatively small scales in which matter
can be treated by field theories in a local inertial frame (LIF). In
a LIF fundamental nongravitational interactions are described by
Lagrangian field theories, and in four spacetime dimensions the mass, $m$,
and spin, $s$, are given by the Casimir invariants of representations of
the local Lorentz group, $P^\mu P_\mu=-m^2c^2$ and $W^\mu W_\mu=s(s+1)m^2c^2
\hbar^2$, where $P^\mu$ is a particle 4--momentum and $W^\mu$ the
Pauli-Lubanski pseudovector. To make a transition to macroscopic scales we
have to take averages of these quantities to obtain a hydrodynamic limit: a
fluid description with an effective energy--momentum tensor, not derivable
from a field Lagrangian.

Following the end of the radiation dominated epoch, on cosmological
scales of averaging, we largely deal with situations in which the
massive matter fields form condensates for which the average
spin is zero, and can be treated as pressureless dust described by
one scalar parameter, the density, $\rh$. Neutrinos, on account of their
ultralight masses, near relativistic speeds, and their interesting property
that mass eigenstates do not coincide with flavour eigenstates, are an
exception to this rule. However, neutrinos are so light that from the
point of view of the cosmological background they can be effectively
treated as a massless species.

Massless particles which are relevant for cosmology will be considered.
Gravitational waves will not be considered here, as their cosmological
contribution at late epochs is negligible. Electromagnetic waves are
considered, as they are the means by which almost all our information
about the universe is transmitted. For the purpose of cosmological averages
it is sufficient that light propagation can be treated in the geometric
optics limit, and that the cosmic radiation background has a perfect fluid
equation of state $P=\frn13\rh c^2$.

The question of the largest scale on which the SEP can apply has not,
to the best of my knowledge, been addressed in a fundamental way. Taken
literally, as soon as we are dealing with scales on which particle
masses must be treated as an integrated density, then the SEP can only
apply if we ignore the gravitational interaction. Of course, in practice,
for many practical purposes we can treat the gravitational interaction
by Newtonian gravity on the scale of the solar system, and the scale
of galaxies. However, the true nature of general relativity is to
replace the Newtonian gravitational force by a dynamical spacetime,
not to replace spacetime by Newtonian gravity in the limit of weak fields.

It must be recalled that whereas Newtonian gravity is a nonlocal
interaction on a static space, general relativity is a local dynamical
theory of spacetime. The curvature of the spacetime background is
not local, and the Newtonian limit picks up the aspect of the nonlocal
curvature of the background generated by slowly moving point particles,
but in full general relativity, changes in the curvature of the background
can only be built up over time by locally propagating processes. This
means that even in the weak--field limit there exist circumstances in
which Newtonian gravity on a static background is not an appropriate
limit. In cosmology the background is not static, and thus clearly the
Newtonian approximation may break down. While much has been achieved
numerically by assuming $N$--body Newtonian interactions on a pre--existing
FLRW background, the universe is clearly inhomogeneous on scales of
at least tens of megaparsecs at the present epoch. To consider the
dynamical gravitational processes which lead to macroscopic variations of
the spacetime background over these scales, we need to go back to
first principles.

\subsection{The cosmological equivalence principle\label{cep}}
My proposed answer to the question of the largest scale on which the SEP can
apply is to
deal with the average effects of density by extending the SEP to potentially
larger scales while removing the time--translation and boost symmetries of
the background as follows:

{\em At any event, always and everywhere, it is possible to choose a suitably
defined spacetime neighbourhood, the cosmological inertial frame, in which
average motions (timelike and null) can be described by geodesics in a
geometry which is Minkowski up to some time-dependent conformal
transformation},
\beq \ds^2\Ns{CIF}=
a^2(\et)\left[-\dd\et^2+\dd r^2+r^2\dOM\right],
\label{cif}\eeq
where $\dOM$ is the metric on a 2--sphere.

The standard SEP is obtained in the limit
that $a(\et)$ is constant, which physically corresponds to virialised
systems that can be effectively thought of as asymptotically flat.
Alternatively, during very short time intervals over which
the time variation of the scale factor can be neglected, the standard
SEP is also retrieved. The idea here is that even when spacetime is
dynamical in cosmology, one can always find a spacetime neighbourhood
whose average volume expansion can be characterised by a metric of the
form (\ref{cif}) for time intervals over which $a(\et)$ varies.
The metric (\ref{cif}) is of course that of the spatially flat FLRW
geometry in conformal coordinates.

The rationale for the statement of the cosmological equivalence principle
(CEP) is twofold. First, if we are to demand a
smooth Newtonian gravitational limit in all circumstances, we have to deal
with the fact that Newtonian gravity deals with just one scalar source,
the density, whereas general relativity is tensorial. This means that
we must be dealing with an average spacetime with symmetries in taking
a Newtonian gravity limit. Since gravitation with matter is dynamical,
it is the symmetries involving time that should be removed from the SEP to
account for the average density of matter. The metric
(\ref{cif}) does this while preserving the isotropy and homogeneity of
space regionally within a cosmological inertial frame (CIF).

\begin{figure}[htb]
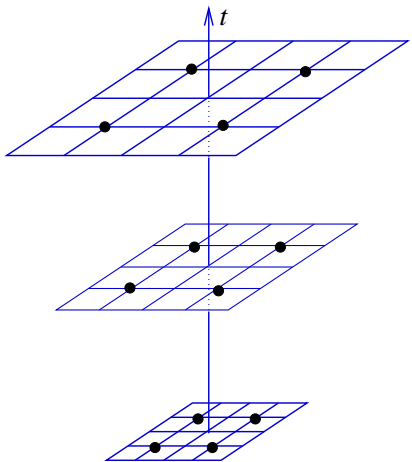

\vbox{\figvolexp
\caption{\label{fig_vol}%
{\sl A set of particles undergoes an isotropic
spatial 3-volume expansion in a spatially flat local region. For the same
initial conditions, provided that we consider time intervals
over which deceleration of the expansion is negligible, it is impossible
to locally distinguish the case of particles at rest in a dynamically
expanding cosmological space from particles moving isotropically in a
static Minkowski space. One spatial dimension is suppressed.}}}
\end{figure}

Second, at the core of the equivalence principle is the notion
that we can always choose a LIF, for example, by specifying Riemann
normal coordinates. However, in the case of the volume--expanding
and contracting motions of the metric (\ref{cif}), as illustrated by
Fig.~\ref{fig_vol}, it is impossible to {\em locally} distinguish between
the case of comoving particles at rest in an expanding metric (\ref{cif})
and the case of particles in motion in the static Minkowski space of the
relevant LIF, a point which has been emphasised by a number of authors
recently from various points of view \cite{doppler}.
On local scales, both yield the Hubble law redshift
$$z\simeq {H_0\ell_r\over c},\qquad H_0=\left.\dot a\over a\right|_{t\X0}$$
where $\ell_r$ is the radial proper distance from an observer at the origin
to a source, and an overdot denotes a derivative with respect to $t$ where
$c\,\dd t=a\,\dd\et$. This is true whether the exact relation, $z+1=a\Z0/a$,
is used or the radial Doppler formula, $z+1=[(c+v)/(c-v)]^{1/2}$, of special
relativity is used, before making a local approximation \cite{foot1}.
Mathematically the equivalence of the two situations might be viewed as
a consequence of $\pt/\pt\et$ being a conformal Killing vector of
(\ref{cif}).

The aim of the CEP is to go beyond the limit of the static special
relativistic LIF, to consider arbitrarily long time intervals over which
the motion of the particles is decelerated, $\ddot a < 0$, by the average
density of matter. As Einstein himself stated \cite{esu}, ``In a consistent
theory of relativity there can be no inertia relatively to `space', but
only an inertia of masses relatively
to one another''. Since the deceleration of the volume expansion is due
to the backreaction of the average density of the matter particles in
defining their average background, the CEP thus represents a refinement
in the definition of inertial frames. To demonstrate this at a conceptual
level, we will first show that for localised regions a suitable equivalent
of decelerated Minkowski particles can always be found for the motion
of a congruence of comoving particles in (\ref{cif}), even for arbitrarily
long time intervals.

\subsection{Thought experiment: The semi-tethered lattice}
Given a spatial point, taken to coincide with the origin
of the coordinates (\ref{cif}), we might be tempted to think that we
can mimic the situation of an arbitrary decelerating expansion of
(\ref{cif}) in Minkowski space by attaching rockets to a set of
test particles initially chosen to be equidistant from the given centre
and with identical radial velocities. The rockets could then be fired
in unison at the same rate to give an inward radial acceleration
while maintaining equidistance of the particles from the centre
and equality of instantaneous radial velocities as seen from
the frame of the central observer with global Minkowski time, $t$.

\begin{figure}[htb]
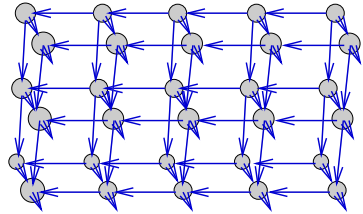

\vbox{\figlattice
\caption{\label{fig_lattice}%
{\sl The semi-tethered lattice. Point particle observers in a homogeneous
spatial lattice, initially expanding with uniform velocity, are
attached to nearest neighbours by strings. Each string is fixed at one end.
The arrowheads denote the free end of each string wound on a spool at a
neighbouring particle, to
which the observers apply brakes in a synchronised fashion according to
a pre-determined plan. The time evolution of the lattice
follows a course similar to that of the spatial grid in Fig.~\ref{fig_vol},
but with deceleration.}}}
\end{figure}

The rocket analogy is flawed, however. To see the flaw, let us consider
a second Minkowski space thought experiment which is directly equivalent
to the cosmological situation of comoving particles moving in the
geometry (\ref{cif}), even in the case of deceleration of the
volume flow. Let us construct what I will call the {\em semi-tethered
lattice} by the following means. Take a lattice of Minkowski
observers, initially moving isotropically away from each nearest neighbour at
uniform initial velocities \cite{foot2}. The lattice observers
are chosen to be equidistant along mutual oriented $\hat x$, $\hat y$ and
$\hat z$ axes. Now suppose that the observers are each attached to
six others by strings of negligible mass and identical tension along the
mutually oriented spatial
axes, as in Fig.~\ref{fig_lattice}. The string in each observer's negative
$\hat x$, $\hat y$ and $\hat z$ directions is held fixed and extends to
the observer's nearest neighbour in those directions. The string
extending towards each nearest neighbour in the positive $\hat x$, $\hat y$
and $\hat z$ directions unreels freely from a spool at the observer's site
on which an arbitrarily long supply of string is wound. The strings initially
unreel at the same uniform rate, representing a ``recession velocity''.
Each observer carries synchronised clocks \cite{foot3},
and at a prearranged local proper time
all observers apply a braking mechanism, the braking mechanisms having
been pre-programmed to deliver the same impulse as a function of local
time.

The semi-tethered lattice is directly analogous to the case of
the decelerating volume expansion of (\ref{cif}) due to some average
homogeneous matter density, because it maintains the homogeneity and
isotropy of space over a region as large as the lattice. In the case of the
firing of rockets, the act of firing a rocket means that each observer with
a rocket feels a net force in a particular direction, while also breaking
the symmetry of the homogeneity of space. In the case
of the semi-tethered lattice work is done in applying the brakes,
and energy can be extracted from this. However, since brakes are
applied in unison (according to local proper time at each lattice site),
there is {\em no net force on any observer in the lattice}. Although the
rate at which the brakes are applied can
be an arbitrary pre-arranged function of local proper time, provided
the braking function is applied uniformly at every lattice site,
the clocks will remain synchronous in the comoving sense, as all observers
have undergone the same relative deceleration.

The semi-tethered lattice is also a useful analogy because the
kinetic energy of the particles is converted directly to heat in
the brakes, which might then be converted to other forms. It is
a direct analogue for the conversion of kinetic energy of particles
in the expansion of the universe to other forms of energy
via gravitational collapse. Apart from the energy of massless and near
massless species released during recombination and earlier phase
transitions, all the useful forms
of energy in the present observable universe have undergone such a process
of transformation: from the kinetic energy of expansion to gravitational
energy, and then to other forms.

The semi-tethered lattice and the CEP might be said to
extend the elevator thought experiments and the 1907 Einstein equivalence
principle \cite{eep} in a natural fashion to the case of a homogeneous
isotropic {\em nonstatic} gravitational field \cite{foot4}.
The important point in the present situation
is that in both the cosmological case and the semi-tethered lattice
analogy the observers feel {\em no net force} from the relative
deceleration, which justifies the description of (\ref{cif})
as a cosmological {\em inertial frame}.

\subsection{Cosmological inertial frames and averaging}
Although (\ref{cif}) is simply the standard spatially flat FLRW
geometry, the important physical difference from the usual treatment of
cosmology is that here it is not to be viewed as a global metric of
spacetime, but as an average cosmological frame over some region. The
FLRW geometries with spatial sections of positive and negative spatial
Gaussian curvature, $k$,
\beq \ds^2\Ns{FLRW}= \ati^2(\eti)\left[-\dd\eti^2
+{\dd\rti^2+\rti^2\dOM\over(1+\frn14k\rti^2)^2}\right],
\label{FLRWgeom}\eeq
can of course also be brought into the form (\ref{cif}) over small
spatial domains as compared to the radius of curvature.
This follows directly from the fact that the FLRW geometries are
conformally flat and coordinates can be found to bring (\ref{FLRWgeom})
into the form \cite{IS}
\beq \ds^2\Ns{FLRW}=
a^2(\et,r)\left[-\dd\et^2+\dd r^2+r^2\dOM\right].
\label{FLRWconf}\eeq

The new physical interpretation is that no single FLRW geometry
(\ref{FLRWgeom}) is to be taken as a global average geometry for the whole
universe for all times. However, with an extension of the SEP to account for
the cosmological average effects of the density of matter, regional
cosmological inertial frames with average geometry (\ref{cif})
can always be found. It should also be emphasised that since these are
average frames, with differing regional scale factors and local coordinates,
no metric of the form (\ref{FLRWgeom}) is substituted into the Einstein
equations and solved. Rather, an appropriate average of the full inhomogeneous
Einstein equations, such as a Buchert average \cite{buch1}, should be applied
to solve for the background average of the inhomogeneous geometry. One
can solve the Buchert equations for a realistic approximation to the observed
universe \cite{sol}, but care must be taken in interpreting the
solution, as we must account for where the observers are within the
inhomogeneous structure when it comes to the relative calibration of their
rods and clocks \cite{clocks}.

In the case of a semi-tethered lattice which is confined to a
finite region, the relative deceleration of the region would give a local
proper time to comoving observers in the lattice different from that of
a global Minkowski observer, even if this observer had a synchronous
clock before the volume deceleration began. By the CEP the average
homogeneous density in different regions likewise sets a standard of local
time for CIFs, and this may vary significantly when regional density
contrasts grow.

Since any CIF (\ref{cif}) is an average, the relationship of the average to
the inhomogeneous geometry needs to be stated. For example, in terms of the
proper time, $c\,\dd t = a\,\dd\et$, of an observer ``comoving'' with the CIF,
the uniform expansion of the CIF should be viewed as an average only:
\beq{\dot a\over a}\equiv\frn13\ave{\th}\Ns{CIF}=H\Ns{CIF},\label{volex}\eeq
where $\th$ is the volume expansion, angle brackets denote the
appropriate average, and an overdot denotes a derivative
with respect to $t$. In relation to the averaging scheme, the specification
of a CIF should incorporate a notion of ``centre--of--mass motion''
in the sense that the variance in the CIF volume expansion
is an average of shear and vorticity fluctuations for which the
net backreaction {\em within} a CIF vanishes. In the notation of
Buchert and Carfora \cite{BC3},
\beq
\de^2H\Ns{CIF}=\frn19\left(\ave{\th^2}\Ns{CIF}-\ave{\th}^2\Ns{CIF}
\right)=\frn13\ave{\si^2}\Ns{CIF},
\label{com}\eeq
in Buchert's scheme with vanishing vorticity, where $\si^2=\frn12\si_{\mu\nu}
\si^{\mu\nu}$ is the scalar shear. The condition (\ref{com}) ensures
that the average kinematic backreaction within a CIF vanishes in
Buchert's scheme. For CIFs containing galaxies, vorticity
will be important, and so Buchert's scheme needs to be generalised to
include vorticity before a precise statement can be made. As the
contributions of vorticity and shear are of opposite sign in the
Raychaudhuri equation their average contributions may even be largely
self--cancelling, for virialised systems at least. However, a precise
formalism including vorticity remains to be developed.
It is possible that ideas could be incorporated from Zalaletdinov's
averaging scheme \cite{Zal}, though the physical content of that scheme
first needs to be more clearly elucidated in a cosmological setting
\cite{foot5}.

For the present paper it is sufficient that the backreaction can be neglected
in characterising the average properties within a CIF. Thus a CIF is
characterised by a single scalar, the volume expansion (\ref{volex}). In
an expanding universe with collapsing substructure, this volume expansion
is best viewed as a macroscopic property of a given CIF, but often
not of its more finely grained subregions. In particular, if a CIF contains
a galaxy or a galaxy cluster it will contain virialised regions and may also
contain collapsing regions. For cosmological averages, when one is interested
in comparing deceleration rates in expanding regions of different density,
the notion of a finite infinity cutoff scale \cite{fit,clocks} is suggested
as a minimum region for a CIF in relevant averages.

\section{Thought experiment: Relative homogeneous isotropic
deceleration\label{gedanken}}

The beauty of the equivalence principle is that it allows one to
quantitatively deduce the order of magnitude of simple effects, such as the
leading order of gravitational redshift \cite{eep}, by simple thought
experiments alone. The basic physical effect that is of interest here -- the
gravitational energy cost of a spatial curvature gradient -- can likewise be
understood by a simple thought experiment.

From the evidence of the cosmic microwave background (CMB) radiation,
we know that, apart from tiny fluctuations of order $\de\rh/\rh\goesas10^{-5}$
in photons and the baryons they couple to, and density fluctuations
perhaps an order of magnitude larger in nonbaryonic dark matter particles,
the observable universe was close to spatially flat,
homogeneous, and isotropic at the epoch of last scattering.
Assuming the Copernican principle, it was sufficient to describe the
universe by a single frame (\ref{cif}) at that epoch. However, since a CIF
is a local regional frame, we should be careful never to construct a CIF with
spatial extent larger than the particle horizon at any epoch. Rather,
it is better physically to think of a class of local CIFs at disjoint spatial
locations at the surface of last scattering, in which there exist
geodesic congruences of observers like those depicted in Fig.~\ref{fig_vol},
which all have an identical uniform expansion away from each other,
on account of the initial uniformity of the Hubble flow at that epoch.

\begin{figure}[htb]
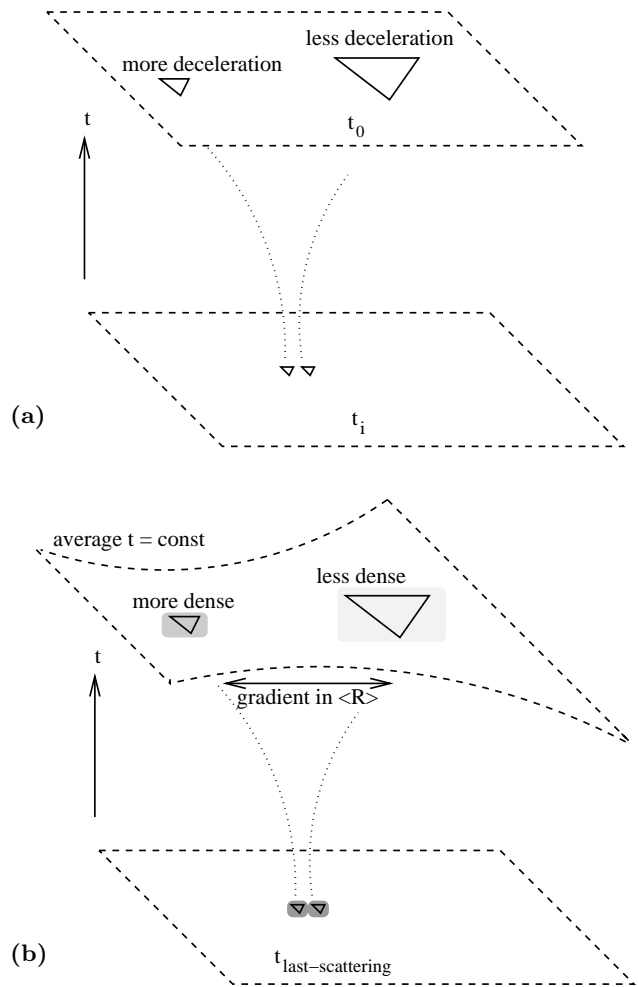

\vbox{\figequivM
\figequivGR
\caption{\label{fig_equiv}%
{\sl Two equivalent situations: {\bf(a)} in Minkowski space observers
in separate semi-tethered lattices, initially expanding at the same
rate, apply brakes homogeneously and isotropically within their respective
regions but at different rates;\break
{\bf(b)} in the universe which is close to
homogeneous and isotropic at last-scattering, observers in separated
regions initially recede from each other at the same rate, but experience
different locally homogeneous isotropic decelerations as local density
contrasts grow. In both cases there is a relative deceleration
of the observer congruences, and those in the region which has decelerated
more will age less.}}}
\end{figure}

Let us first analogously consider two sets of disjoint semi-tethered
lattices, with identical initial local expansion velocities, in a
static background Minkowski space. [See Fig.~\ref{fig_equiv}(a).]
The observers in the first congruence apply brakes in unison to decelerate
homogeneously and isotropically, with inward 4--acceleration of magnitude,
$\al\Z1(\ta\Z1)$, as measured by the force applied to the brakes in the
frame of any of the observers, where $\ta\Z1$ is the proper time measured
by any of them. From the viewpoint of a global Minkowski observer, members
of the congruence will agree on their measurements of time,
$\ta\Z1$, though this time of course differs from the global Minkowski
observer's time, $t$.

Now take a second semi-tethered lattice, with the same initial expansion
speed, where brakes are applied with a force corresponding to a 4--acceleration
of magnitude $\al\Z2(\ta\Z2)$.
At any global Minkowski time, $t$, we will assume that
when transformed from their proper frames to that of the
global Minkowski observer, at each time step
$\al\Z1(t)>\al\Z2(t)>0$. It is then the case that the members of the
first congruence decelerate more than the members of the second congruence,
and at any time $t$ the proper times satisfy $\ta\Z1<\ta\Z2$. The members
of the first congruence age less quickly than members of the second
congruence.

By the CEP, the case of volume
expansion of two regions of different average density at late times
is entirely analogous. The fact that we are able to apply the equivalence
of the two circumstances rests on the fact that the expansion
of the universe was extremely uniform at the time of last scattering,
by the evidence of the CMB. At this epoch all regions effectively
have the same density -- apart from negligible fluctuations -- and
the same uniform Hubble flow. At late epochs, suppose that in the frame of
any average cosmological observer there are regions of different
density which have decelerated by different amounts by a given time, $t$,
according to that observer. Then, by the CEP the local proper time of the
isotropic observers in the denser region, which has decelerated more,
will be less than that of the equivalent observers in the less
dense region which has decelerated less. [See Fig.~\ref{fig_equiv}(b).]
Consequently the {\em proper time of the observers in the more
dense CIF will be less than that of those in the less dense CIF}, by
equivalence of the two situations.

The fact that a global Minkowski observer does not exist in the second
case does not invalidate the argument. The global Minkowski time is
just a coordinate label, and in the cosmological case the only restriction
on the $t=$const slices is that {\em the expansion of both average congruences
must remain homogeneous and isotropic in local regions of different average
density} in the global $t=$const slice. Of course, we need to be careful to
patch together different CIFs continuously to specify the slice, as we will
further discuss in Sec.\ \ref{frame}. In this way the equivalence to the
Minkowski space case is maintained. Thus in the cosmological case, provided
that we refer to {\em local} homogeneous isotropic expansion in different
regions on any average $t=$const slice, (where $t$ is some coordinate
label), then if such regions {\em are still expanding} and have
a significant density contrast, we can expect a significant clock rate
variance.

This equivalence directly establishes the idea of a {\em gravitational
energy cost for a spatial curvature gradient}, since the existence
of expanding regions of different density within an average $t=$const
slice implies a gradient in the average Ricci scalar curvature, $\Rav$,
on one hand, while the fact that the local proper time varies
on account of the relative deceleration implies a gradient in gravitational
energy on the other.

\subsection{Order of magnitude estimate of clock rate variance\label{order}}

An order of magnitude estimate of present epoch clock rate variances
due to gravitational energy gradients induced by relative volume
deceleration of the background can now be made by directly
using observationally measured density contrasts. Although a CIF
is a frame (\ref{cif}) within which backreaction can be neglected,
to determine its scale factor over long periods of time one
must consider the evolution of the universe within which the CIF is
embedded. Such evolution will, in general, include the effects of
backreaction. However, if the backreaction is small,
an order of magnitude estimate of the clock rate variance can
be made assuming that regions with observed strong density contrasts
evolve independently by solutions of the local Friedmann equation for
regions of different density.
There will be a relative deceleration of the local background
of such regions, which via equivalence to the Minkowski space
semi-tethered lattices, will accumulate clock rate differences.

Galaxies formed from perturbations which were greater than critical
density and if space is negatively curved on average, they must always be
bounded by a region which is spatially flat. These on--average spatially
flat locally expanding bounding regions are called {\em finite infinity}
regions \cite{clocks,fit}, and a union of such regions is called a {\em wall}.
Since they are spatially flat, neglecting backreaction,
the evolution of the wall CIFs can be approximated by spatially flat
Einstein--de Sitter regions with local scale factor
$\aw=a_i\left(\frn32H_i\tw\right)^{2/3}$, where $H_i$
is the common initial Hubble parameter at last scattering, and $a_i$
is a constant. On the other, hand CIFs within voids can be approximated
by portions of spatially open FLRW solutions, given parametrically by
\bea
\av&=&{a_i\omi\over2(1-\omi)}(\cosh\eti-1)\,,\label{scalev}\\
H_i\tv&=&{\omi\over2(1-\omi)^{3/2}}(\sinh\eti-\eti)\,,
\label{agev}\eea
where $c\,\dd\tv=\av\,\dd\eti$, $\omi=1-\ep_i$ is an initial density
parameter at last scattering, $\ep_i\ll1$ and $a_i$ is a constant. Using
(\ref{scalev}) and (\ref{agev}) in the Friedmann equation, one obtains a
standard parametric relation for the density parameter,
\beq
\OM(\eti)={2(\cosh\eti-1)\over\sinh^2\eti}\,,
\label{oms}\eeq
which is to be viewed here as a regional parameter in a CIF inside a
void.

We now follow the analysis of Ref.\ \cite{paper0}, where the author's
proposal was first advanced. There an attempt
was made to estimate cosmological parameters by making the approximation
that the evolution of the observable universe was entirely due to the
voids. In fact, backreaction between the walls and voids must be
included to obtain more reliable estimates of cosmological parameters
\cite{clocks,LNW}. However, if we confine attention to small regional CIFs,
then the argument of Ref.\ \cite{paper0} can give an estimate of clock
rate variance from {\em observed} density contrasts.
In particular, since the critical density
also defines the Einstein--de Sitter standard of time of the wall CIFs,
it also follows that
\beq
\OM=\omi\left(\aw\over\av\right)^3=
{18{H_i}^2(1-\omi)^3\tw^2\over{\omi}^2(\cosh\eti-1)^3}\,.
\label{bom}\eeq
Combining (\ref{oms}) and (\ref{bom}) we find
\beq
H_i\tw={\omi(\cosh\eti-1)^2\over3(1-\omi)^{3/2}\sinh\eti}\,.
\label{agew}\eeq
Differentiating both (\ref{agev}) and (\ref{agew}) we find a relative
clock rate, which we will call the {\em lapse function}, given by
\beq
\ga(\eti)\equiv\Deriv{\dd}\tw{\tv}={3(\cosh\eti+1)\over
2(\cosh\eti+2)}\,. \label{gam1}\eeq

A relative clock rate variance due to the relative volume deceleration
between CIFs in walls and voids can now be estimated
since $\OM=1+\deti$, where $\deti$ is the density contrast, and inverting
(\ref{oms}) we have
\beq\cosh\eti={2-\OM\over\OM}={1-\deti\over1+\deti}.
\label{contrast}\eeq
Hoyle and Vogeley \cite{HV} estimate that 40\%--50\% of the present--epoch
universe is in voids of diameter about $30h^{-1}$Mpc, with
the statistics summarised in Table \ref{tab_void}. The density contrasts
quoted are an average, and are of greater magnitude in the centres
of the voids which are extremely empty. Taking these values as
indicative, if we assume $\deti=-0.9$ then (\ref{gam1}) and (\ref{contrast})
give $\ga=1.42$, while if we assume $\deti=-0.95$ then $\ga=1.46$. For
larger density contrasts, in the centre of the void the lapse would approach
the limiting value $\ga\to1.5$, which represents the relative local
expansion rates of an empty Milne universe \cite{Milne} to an
Einstein--de Sitter one.

These clock rate variances of 42\%--46\% are large, and counter intuitive
given we usually encounter large clock rate differences only for
large local boosts or for density contrasts from extremely compact
sources in static backgrounds, such as black holes. However, the effect
described is neither of these familiar situations, as the {\em background
is not static}. Rather, it is the
clock rate variance due to the cumulative effect of a very small relative
deceleration of the background. The above variances are simply those
demanded by the CEP taking present epoch density contrasts observed in the
actual universe.
\begin{table}[hb]
\begin{center}
\begin{tabular}{l|c|c}
\hline
Survey &\quad Void diameter &\quad Density contrast \\
\hline
PSCz & $(29.8\pm3.5) h^{-1}$Mpc & $\deti=-0.92\pm0.03$\\
UZC & $(29.2\pm2.7) h^{-1}$Mpc & $\deti=-0.96\pm0.01$\\
2dF NGP & $(29.8\pm5.3) h^{-1}$Mpc & $\deti=-0.94\pm0.02$\\
2dF SGP & $(31.2\pm5.3) h^{-1}$Mpc & $\deti=-0.94\pm0.02$\\
\hline
\end{tabular}
\caption{Dominant void statistics in the Point Source Catalogue Survey
(PSCz), the Updated Zwicky Catalogue (UZC), and the 2 degree Field
Survey (2dF) North Galactic Pole (NGP) and South Galactic Pole (SGP),
from Refs.\ \cite{HV}.\label{tab_void}}
\end{center}
\end{table}

Taking backreaction into account to determine
a full present epoch average \cite{clocks,sol}, the
present epoch lapse difference between walls and a volume--average location
in a void based on luminosity distance data fits is $\gb=1.38^{+0.06}_{-0.05}$
\cite{LNW}. This is an average value; the relative lapse would be
larger in the void centres. Thus the estimates made without backreaction
are reasonably accurate, showing that the effect is not a direct consequence
of backreaction in the evolution equations but rather of relative volume
deceleration alone. From (\ref{gam1}) and (\ref{contrast}) without
backreaction the density contrast estimate for a local CIF at the
volume--average position is $\deti=-0.83$ for
a lapse of $\ga=1.38$. If such a clock variance were produced by a uniform
acceleration in Minkowski space, a simple calculation shows it would require
a tiny relative acceleration of order $5.5\times10^{-10}$ms$^{-2}$
over the lifetime of the universe. Of course, such a relative acceleration is
not uniform: a better estimate is presented in Sec.\ \ref{alpha}.

The argument above relies on it being possible to choose a locally uniform
Hubble flow gauge, as will be discussed in the next section. Such a gauge can
be maintained outside finite infinity regions, but not within them where
collapsing regions are located, and where vorticity and tidal torques
become important. Thus there are no obvious inferences analogous to those
above that can be made with respect to bound structures from
observations of the magnitude of their positive density contrasts.

\section{Average frames\label{frame}}
\subsection{The cosmic rest frame\label{rest}}

In taking cosmological averages with inhomogeneous structures, the question
arises as to which average frames have the most utility. One must generally
make a choice of gauge in specifying such frames. One way of viewing the
SEP is that we can always set the first derivatives of the metric to zero
near a point. In particular, the volume expansion $\th$, which involves
first derivatives of the metric, can always be set to zero. The CEP extends
this by the statement that in the dynamical situation a gauge can be
chosen in which the volume expansion in a CIF is spatially uniform, but
varying with time.

As was pointed out in Sec.\ \ref{gedanken} in order to compare CIFs in
regions of different densities, we need to specify suitable spacelike
slices \cite{foot6} by patching together different CIFs
in a continuous manner.
Operationally, the way to do this is first to choose an orientation of
the 4--velocity fields, $\pt/\pt\ta_\I$, of comoving observers in a CIF
{\em such that the CMB radiation is isotropic} in each frame.
In terms of the local proper time, $\ta_\I$, of such observers the metric
(\ref{cif}) is rewritten
\beq \ds^2\Ns{CIF}=
-c^2\dd\ta_\I^2+a_\I^2(\ta_\I)\left[\dd r_\I^2+r_\I^2\dOM\right],
\label{cif2}\eeq
where the index $\scriptstyle I$ runs over CIFs.
Secondly the spacelike slice is specified by the demand that the {\em locally
measured value} of the volume expansion remains uniform as one moves from
the patch of one CIF to the next. In other words, the ``local Hubble flow''
remains uniform in this gauge even though the proper lengths and proper
time scales will change as one moves between CIFs of different density.
As discussed in Ref.\ \cite{clocks}, although the proper volume of
voids increases faster than that of wall regions, this is compensated
for {\em locally} by the faster relative rate of the void clocks. Relative
to any one set of clocks, such as our own, it
will always appear that voids expand faster than walls. So the average
Hubble flow over both walls and voids -- by one set of clocks -- will
generally differ from the underlying uniform flow. Its value is a choice of
gauge depending on the choice of fiducial observer.

The uniform Hubble flow slices defined in this manner constitute the
{\em cosmic rest frame}: surfaces within which the CMB is isotropic,
even though the mean value of the CMB temperature, and angular scale
of CMB anisotropies, will vary from
point to point as spatial volumes vary in relation to proper radius
with changes in spatial curvature. The proper times of CIF observers
within the slice will also vary. We can choose the clocks of a canonical
set of observers in expanding regions with the same average local
density to label the slices, provided we realise that this time
labelling is only a proper time in particular locations on the slice
and is just a coordinate label elsewhere.

The uniform Hubble flow gauge is one of the standard gauges of
perturbation theory in FLRW models \cite{Bard}, and has been further
refined with the addition of a minimal shift distortion condition by
Bi\v{c}ak, Katz and Lynden-Bell \cite{BKL}. These authors recognise
the resulting ``Mach 1 gauge'' as one of three possible gauges which best
incorporates Mach's principle, and within which there is a minimal amount
of residual gauge freedom. Bi\v{c}ak, Katz and Lynden-Bell work within the
framework of perturbations of a global FLRW geometry. The viewpoint of the
present paper is that there is no single global FLRW geometry. However,
on account of the CEP the spatially flat and negatively curved
FLRW geometries can be considered as local regional geometries
on spacelike slices as the density varies relative to the critical
density. Thus a modification of the formalism of Bi\v{c}ak, Katz and
Lynden-Bell may be an appropriate starting point to deal with an
averaging formalism for the full nonlinear inhomogeneous problem,
as an alternative to Buchert's scheme.

\subsection{Buchert averaging}

Buchert's averaging scheme \cite{buch1} is based on the starting point
that, in the case of an energy--momentum tensor for irrotational dust
particles in the presence of inhomogeneities, one can choose
Gaussian normal coordinates
\beq
\ds^2=-c^2\dd t^2+\gd_{ij}\dd x^i\dd x^j,
\label{gnc}\eeq
comoving with the dust. The scalar
density appearing in the energy--momentum tensor,
$$T^{\mu\nu}=\rh c^2\bn^{\mu}\bn^{\nu}$$
where $\bn^\mu=\Deriv{\dd}t{X^\mu}$, then represents the rest mass density of
the dust, and one averages over spatial slices of constant $t$ orthogonal
to the flow, over regions which conserve the rest mass of a portion of
the fluid in a domain, $\DD$, with continuity equation
\beq
\pt_t\ave\rh+{\dot\ab\over\ab}\ave\rh=0,
\label{cont1}\eeq
where $\ab(t)\equiv\left[\Vav(t)/\Vav(t\Z0)\right]^{1/3}$ with
$\Vav(t)\equiv\int_\DD\dd^3x\sqrt{\det\gd}$. Here angle brackets denote the
spatial volume average of a quantity, so that
$\Rav\equiv\left(\int_\DD\dd^3x\sqrt{\det\gd}\,{\cal R}(t,\bx)\right)/\Vav(t)$
is the average spatial curvature, for example. The Buchert
equations consist of (\ref{cont1}) and
\bea
&&\dsp{3\dot\ab^2\over\ab^2}=8\pi G\ave\rh-\half c^2\Rav-\half\QQ,
\label{buche1}\\
&&{\ddot\ab\over\ab}=-4\pi G\ave\rh+\QQ,\label{buche2}\\
&&\pt_t\left(\ab^6\QQ\right)+\ab^4 c^2\pt_t\left(\ab^2\Rav\right)=0,
\label{buche3}\eea
where
$\QQ=\frn23\left(\langle\th^2\rangle-\langle\th\rangle^2\right)-
2\langle\si^2\rangle$,
is the kinematic backreaction. Equation (\ref{buche3}) is an integrability
condition which ensures closure of the other equations.

Since the backreaction term $\QQ$ includes variance in volume expansion,
and this is to be evaluated on a constant $t$ slice, it is clear that
as compared to the cosmic rest frame of Sec.\ \ref{rest}, different
physical premises underlie the interpretation implicitly assumed by
Buchert's scheme. My approach is therefore different from other approaches
to cosmological building that have been adopted in the context of
Buchert averaging \cite{morph,Rasanen,LS,PS2}.
The differences may be understood from the fact that in talking about the
``rest mass density of the dust'' one is actually dealing
with a concept which depends on the manner in which dust ``particles'' are
coarse grained. Since all forms of energy have a
rest mass equivalent, the kinetic energy of particles within a dust
particle is included as rest mass. Similarly, since Ricci
curvature affects spatial volumes relative to their diameter, the concept
of a rest density depends on the scale of coarse graining relative
to the curvature scale.

In general, the notion of ``comoving with the dust'' implicit in Buchert's
scheme can be very distinct from ``comoving with the background'', although
the notions coincide for FLRW models. This is well illustrated by the exact
spherically symmetric Lema\^{\i}tre--Tolman--Bondi (LTB) models \cite{LTB}
for pressureless dust, with a prescribed inhomogeneous density $\rho(t,r)$.
These can be written in Gaussian normal form (\ref{gnc}), making it
straightforward to compute a Buchert average \cite{BA,Sussman} \cite{foot7}.
At fixed comoving proper time, $t$, as the radial coordinate
$r$ varies the LTB dust shells have different densities, different spatial
curvature, nonzero shear, and, in general, observers at $r>0$ would not
expect to see an isotropic CMB. Since the solution is completely dynamical,
there is no average homogeneous isotropic background with
respect to which one could be comoving, unless one puts in such a background
by hand by making the model asymptotic to an FLRW model at large $r$.

With respect to fixed FLRW backgrounds, an alternative simple way to treat
spherical inhomogeneities is by the spherical top hat model, using concentric
spherical shells \cite{GG,SW}. In the case of a void in a background
Einstein--de Sitter universe, for example, a spherical underdense
shell will acquire a peculiar velocity with respect to the background
which tends to 50\% of the background Hubble rate at late times \cite{foot8}.
One can account for the kinetic energy of the shell, but in view of the large
peculiar velocity of the shell there is a limit to the extent to which it
can be considered comoving with respect to the background with a
synchronous clock.

Once one averages on the scale of statistical homogeneity, as in (\ref{cont1}),
one wants to have a sense of ``comoving with the background''; i.e.,
different observers in different averaging cells should have a notion of
determining the same average density at the same cosmological epoch,
and one should be able to talk about motion with respect to canonically
defined observers.

In general, when the background is only {\em statistically} homogeneous and
isotropic, there is an ambiguity in distinguishing between motion
of the background and motion with respect to the background. The spirit
of the CEP is that in the case of volume expansion, this is because there
is a fundamental indistinguishability: the Hubble parameter is a gauge
choice. The definition of the uniform Hubble expansion gauge of Sec.\
\ref{rest} makes an unambiguous separation of the ``kinetic energy of
the volume expansion of space'' for regions which are {\em locally
spatially flat} from other forms of energy. In the standard
interpretation of Buchert averaging, depending on the choice of averaging
volume and the manner in which one coarse grains over dust cells,
the kinetic energy of the volume expansion of space can be intermingled
with other forms of energy.

The view that I adopt is that one can either use Buchert averaging in
defining CIFs on small scales by the requirement that the kinematic
backreaction can be neglected, as in (\ref{com}), or alternatively,
on dust cells of at least the scale of statistical homogeneity,
of about $100h^{-1}$Mpc, at which scale the volume expansion is
statistically uniform. In both cases the
time parameter and $\Rav$ can be regarded as relevant
parameters which describe the {\em collective} degrees of freedom of
the cell. However, they should not be regarded as representative
for observers in finer
partitions within the cell. For CIFs containing galaxies with black
holes this imperative is obvious. In the case of statistically
homogeneous cells in cosmology, the case has been less obvious,
but I believe it is equally imperative on account of the density
contrasts that are observed between voids and walls below the
scale of homogeneity, together with the arguments of
Sec.\ \ref{gedanken}.

I will take the view that the Buchert time parameter
is the relevant one for an observer at a volume--average position
in a statistically homogeneous cell. As the present universe is
dominated by voids, this will be in a void but not at the void
centres. Kinematic backreaction between
voids and walls from the volume--average perspective
must be included in determining the average evolution.
Einstein's equations are causal and depend on all events within the
past light cone. Thus some sort of spatial averaging, such as Buchert
averaging, is required to determine cosmic evolution.
Buchert's equations should thus be viewed as evolution equations.

Observations are made on null cones, however. Thus a Buchert average is
not the one we perform operationally in determining cosmic averages.
Instead, a radial null geodesic average
of a solution to the Buchert equations, combined with an operational
identification of relevant classes of observers within the
inhomogeneous structure, is required to make comparisons with
observations. This is the approach adopted in Refs. \cite{clocks,sol,LNW}.

\section{Relative deceleration of the background and the weak--field
limit\label{alpha}}

In Refs.\ \cite{clocks,sol} a model universe was constructed based on
a regional division of cells of average homogeneity into voids, and the
bubble walls that surround them, assuming that backreaction within the
walls and voids can be neglected, but not in the combined average
\cite{foot9,foot10}.
Space within the walls is assumed to
be spatially flat on average, and space within the voids is negatively
curved. Technically a ``wall'' is understood to be a union of connected
finite infinity regions \cite{clocks}, namely
CIFs in regions of average critical density. Observationally, the
walls would include all morphological types of extended structures
containing galaxy clusters:
namely ``sheets'', ``filaments'' and ``knots'' \cite{forero}.

Qualitatively the regional division into voids and walls may well be
a consequence of the evolution of the scales with statistical excesses
of power in the primordial angular power spectrum. The $100h^{-1}$ Mpc scale
of statistical homogeneity would correspond to the first Doppler peak
(BAO scale): on account of the finite sound speed in the primordial plasma
no structures in excess of this scale are expected statistically, with the
exception of those random structures that arise by percolation. The scale
of the $30h^{-1}$ Mpc dominant voids would correspond to the evolution
of the second Doppler peak, namely, the first rarefaction peak, which is well
within the nonlinear regime of structure formation. The third Doppler
peak, which is the first compression peak within the nonlinear regime,
would give the scale of the largest bound structures that have broken
from the Hubble flow, namely, clusters of galaxies. Finite infinity represents
a demarcation scale of on--average spatially flat regions between clusters
of galaxies and voids. The fourth Doppler peak, the next rarefaction peak,
may possibly give an independent scale corresponding to minivoids. In the
two--scale approximation \cite{foot11}
of Refs.\ \cite{clocks,sol}, we assume that the average curvature of minivoids
and dominant voids is the same. A further refinement might separate these
scales. Ultimately these qualitative speculations about the correspondence
of the Doppler peaks to the observed scales of present epoch structures
should be verified from a numerical model of structure formation.

An exact solution to the Buchert evolution equations of the
two scales was found \cite{sol}, which by a Bayesian analysis fits the
Riess07 gold supernovae data set \cite{Riess07} at a level which is
statistically indistinguishable from the standard spatially flat
$\Lambda$CDM model \cite{LNW}. The same best-fit parameters also fit
the angular scale of the sound horizon seen in CMB data, and the effective
comoving baryon acoustic oscillation scale seen in angular diameter
tests of galaxy clustering statistics \cite{clocks,LNW}.

Those sceptical of the proposal sometimes question how the relatively
large present lapse of $\gb=1.38$ between wall observers and the volume
average can have possibly arisen. As pointed out already in Sec.\ \ref{order}
a relatively small relative deceleration of the background for the
lifetime of the universe is sufficient to achieve this. Since the
accumulated average lapse function $\gb$ is not uniform in time,
the equivalent relative deceleration of the background is not uniform.
In this section we will estimate the relative deceleration which would
produce the tracker solution mean lapse function \cite{sol}.

In principle, one is trying to compare the relative deceleration
of regions of different density as a function of the expansion
age of the universe of some fiducial observer such as ourselves. Ideally,
if sources did exist freely falling in voids unbound to condensed structures,
then by the assumptions of Ref.\ \cite{clocks}
they would have different redshifts to sources in bound structures
that coexist ``at the same epoch'' on account of the accumulated
gravitational energy differences. We would have
\beq
1+z\ns{w}={(-\bk\cd\bU\ns{w})\over(-\bk\cd\bU\ns{obs})}
\eeq
for the redshift of a wall source with 4-velocity $\bU\ns{w}$ as seen
by an observer with 4-velocity $\bU\ns{obs}$, where $\bk$ is the 4-velocity
of a radial null geodesic. Similarly
\beq
1+z\ns{v}={(-\bk\cd\bU\ns{v})\over(-\bk\cd\bU\ns{obs})}
\eeq
where $\bU\ns{v}$ is a volume--average void source, and $z\ns{v}\ne z\ns{w}$.
In general \cite{clocks}
\beq
1+z\ns{w}=(1+z\ns{v})/\gb
\label{relapse}\eeq
where similarly to (\ref{gam1}) $\dd t=\gb\,\dd\tw$ represents the
relative clock rates of volume--average observers to wall observers
at the epoch of emission.
Unfortunately, we have a mass biased view of the universe and only
observe sources in bound systems in regions which are locally spatially
flat on average; otherwise the relative blueshift $-1+\gb^{-1}$ of a
volume--average void clock relative to a wall one would have an obvious
observational signature nearby.

To determine the relative deceleration, we will invoke the CEP by
considering the Minkowski space equivalent
semi-tethered lattice analogy of Fig.\ \ref{fig_equiv}. Ideally,
we should have to calculate the difference in the rate of extraction
of energy in applying brakes at different rates, to represent regions of
different density in the actual universe. However, this should be
equivalent to asking what relative
volume deceleration would be required to produce $\gb(\tw)$ at any
epoch, if $\gb(\tw)$ is treated as a Lorentz gamma-factor in special
relativity, beginning from two regions with the same initial
expansion velocity (which is the situation at last scattering).
In relation to (\ref{relapse}), this is equivalent to
treating the blueshift $-1+\gb^{-1}$ of voids relative
to walls as if it were a standard transverse Doppler shift in special
relativity. Since we are dealing with an isotropic volume deceleration
there is no directional significance associated with a ``direction
of motion'' in the special relativistic analogy.

We assume equivalence to the special relativistic 4--acceleration
$\alB=\Deriv{\dd}\ta\bU$, where $U^\mu=\ga(c,v^i)$, which has a magnitude
\cite{Rindler}
\beq
{\al\over c}={1\over\sqrt{\ga^2-1}}\Deriv{\dd}\ta\ga=\Der\dd t
\sqrt{\ga^2-1},
\eeq
with $\dd t=\ga\,\dd\ta$. Of course, in the present case we are not
really dealing with a proper acceleration in a single direction, as the
appropriate analogy is that of two semi-tethered lattices in which
all directions contribute. Assuming equivalence of the situations
the relative acceleration of the background here has a magnitude
\beq
{\al\over c}=\Der\dd t\sqrt{\gb^2-1}={\gb(\gb\bH-H)\over\sqrt{\gb^2-1}}
\label{a0}\eeq
where, following the notation of Refs.\ \cite{clocks,sol}, $t$ is the
time parameter of the volume--average observer in a void, and
$\dd t=\gb\,\dd\tw$, where $\tw$ is the time for an observer at finite
infinity in a wall, which will be close to the time in an average galaxy.
Furthermore, $\bH$ is the bare or locally measured Hubble parameter,
while on account of eq.\ (42) of Ref.\ \cite{clocks}, $H=\gb\bH-\dot{\gb}$,
is the dressed Hubble parameter obtained by averaging over both walls and
voids on the past light cone.

Using the tracker solution of Ref.\ \cite{sol},
\bea
\gb&=&\frn32t\bH(t)\\ &=&1+\half\fv(t)\\
&=&{9\fvn\Hb t+2(1-\fvn)(2+\fvn)\over2\left[3\fvn\Hb t+(1-\fvn)(2+\fvn)
\right]}\,,\label{gam2}
\eea
where $\Hb$ is the present epoch value of the bare Hubble parameter $\bH(t)$,
$\fv(t)$ is the void fraction and $\fvn$ its present epoch value. The void
fraction here is that of all voids, including minivoids \cite{minivoids},
and not just the dominant voids of Table \ref{tab_void}. The
dressed Hubble parameter satisfies
\bea
H&=&{2\over3t}+{\fv(t)[4\fv(t)+1]\over6t}\nonumber\\
&=&\bH(t)+{\fv(t)[4\fv(t)-1]\over6t}\,,
\eea
while the time parameter $\tw$ of wall observers is related to
that of volume--average ones by
\beq
\tw=\frn23t+{4\OmMn\over27\fvn\Hb}\ln\left(1+{9\fvn\Hb t
\over4\OmMn}\right)\,, \label{tsol}
\eeq
where $\OmMn=\frn12(1-\fvn)(2+\fvn)$ is the present epoch dressed matter
density.
From (\ref{a0})--(\ref{gam2}) we find
\beq
{\al\over c}={3(1-\fvn)(2+\fvn)\fv(t)\bH(t)\over2\sqrt{3\fvn\Hb t
\left[15\fvn\Hb t+4(1-\fvn)(2+\fvn)\right]}}\,.\label{a3}
\eeq

\subsection{Estimate of relative deceleration scale}
Using the estimates of Ref.\ \cite{LNW}, at the present epoch the
void fraction is $\fvn=0.76^{+0.12}_{-0.09}$, while the dressed Hubble
constant is $H\Z0=61.7^{+1.2}_{-1.1}\kmsMpc$, where the uncertainties are
1$\si$ uncertainties from a fit to the Riess07 gold data set \cite{Riess07}.
The bare Hubble constant is then $\Hb=48.2^{+2.0}_{-2.4}\kmsMpc$. From these
values the present epoch magnitude of the relative deceleration (\ref{a3})
is $\al\Z0=6.7^{+2.4}_{-3.4}\times10^{-11}$ms$^{-2}$. Furthermore, since
$\al$ is a time--varying quantity, its best--fit value is plotted
as a function of redshift for recent epochs in Fig.~\ref{fig_accel1}.
\begin{figure}[htb]
\vbox{\figaccela\figaccelb
\caption{\label{fig_accel1}%
{\sl The magnitude of the relative deceleration scale, $\al$, as a function
of redshift: {\bf(a)} for redshifts $z<0.25$; {\bf(b)} for redshifts $z<2$.
The central curve shows the value for the best--fit
parameters $\fvn=0.76$, $H\Z0=61.7\kmsMpc$ ($\Hb=48.2\kmsMpc$) from
Ref.\ \cite{LNW}. The dashed curves show the corresponding results
with $1\si$ uncertainties, which are largely due to the large uncertainty
on $\fvn$, which is not tightly constrained by supernovae data. The
upper dashed curve corresponds to $\fvn=0.67$ and the lower dashed curve to
$\fvn=0.88$. In panel {\bf(a)} the horizontal dotted lines indicate the
bounds of the empirical acceleration scale of MOND when normalised for
$H\Z0=61.7^{+1.2}_{-1.1}\kmsMpc$.}}}
\end{figure}

It is interesting to note that over the range of redshifts $z\lsim0.25$
in Fig.~\ref{fig_accel1}, the curve for the
best--fit parameters with present epoch void fraction $\fvn=0.76$
precisely traverses a range of values for $\al$ that has been used for the
empirical acceleration scale of the modified Newtonian dynamics (MOND)
scenario \cite{mond}. In particular, using a range of recently quoted values
\cite{McGaugh}, the MOND acceleration scale is
$\al\ns{mond}=1.2_{-0.2}^{+0.3}\times10^{-10}h_{75}^2$ms$^{-2}$,
where $h_{75}=H\Z0/(75\kmsMpc)$. Using our best fit
$H\Z0=61.7^{+1.2}_{-1.1}\kmsMpc$, this range
corresponds to $\al\ns{mond}=8.1^{+2.5}_{-1.6}\times10^{-11}$ms$^{-2}$.

\begin{figure}[htb]
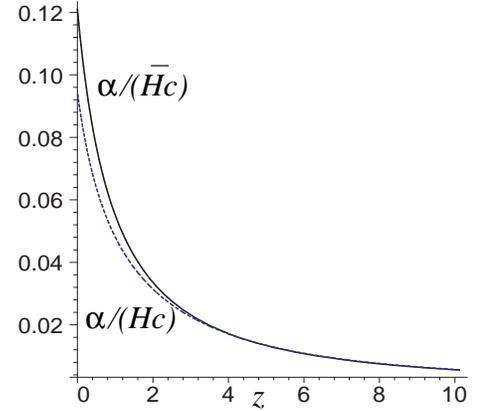

\vbox{\figaccelHc
\caption{\label{fig_accelHc}%
{\sl The magnitude of the dimensionless ratios $\al/(c\bH)$ (solid curve) and
$\al/(cH)$ (dashed curve), where $\al(z)$, $\bH(z)$, and
$H(z)$ are varied with redshift, for best--fit values of $H\Z0$, $\fvn$.}}}
\end{figure}

The fact that $\al$ is larger at higher redshifts reflects the
property that it scales in proportion to
the bare Hubble parameter, $\bH(t)$, as can be seen from (\ref{a3}).
Both the bare Hubble parameter and dressed Hubble parameter are of course
larger at earlier epochs at higher redshift, since the universe is
always decelerating in the present model. (As demonstrated in Refs.\
\cite{clocks,sol} cosmic acceleration is a purely apparent effect related
to clock rate variance.)
From (\ref{a3}) one sees that $\al$ also scales in
proportion to $\fv(t)$ which is smaller at earlier times, contributing
a term in competition with both the term $\bH(t)$ in the numerator of
(\ref{a3}) and with terms in the denominator. As a proportion
of the Hubble flow at any epoch, the relative deceleration is {\em suppressed}
at higher redshifts, as is shown in Fig.~\ref{fig_accelHc}, where the
dimensionless ratios $\al/(\bH c)$ and $\al/(H c)$ with respect to
the bare and dressed Hubble parameters are plotted. At last
scattering, $z\simeq1100$, $\al/(\bH c)\goesas 6\times10^{-6}$.
As $t\to0$, $\al/(\bH c)\propto t^{1/2}$.

The object of estimating the relative deceleration scale was to determine
whether its magnitude is physically
acceptable. This is the case. Although the physical magnitude of
the relative deceleration is larger at earlier epochs, over most of the
history of the universe $\al$ is of the order of $10^{-10}$ms$^{-2}$,
which is tiny.
From Fig.~\ref{fig_accel1}, at $z=2$ we have $\al\simeq4.5\times10^{-10}
$ms$^{-2}$, which corresponds to an expansion age of $4$Gyr,
27\% of the current age of the universe in wall time. By comparison, the
Pioneer anomaly in the solar system \cite{Pioneer} occurs at an
acceleration scale of $(8.74\pm1.33)\times10^{-10}$ms$^{-2}$ \cite{foot12},
a value which $\al$ attains only at a redshift of 3.85 equivalent
to an expansion age of $2.1$Gyr in wall time. Furthermore, the relative
deceleration scale should largely affect dynamics in the transition zones
between walls and voids. At earlier epochs the void fraction is less: at
$z=2$ it is 44\%, and at $z=3.85$ it is 28\%. At $z=10$, when
$\al\simeq2\times10^{-9}$ms$^{-2}$, it is 10\%.

In the absence of an exact timelike symmetry of the background there is
no obvious solution to the problem of how to keep two clocks synchronised
in general relativity. The CEP proposes a solution to this conundrum: the
evolving average density provides the relevant regional clock. Even
though we are talking about weak fields in cosmology, and small relative
decelerations between expanding regions of different densities, the fact
that the relative decelerations are integrated over the lifetime of the
universe means that the cumulative clock variance can be large for the
density contrasts that are observed.

\subsection{Modified Newtonian dynamics?}
The fact that the present epoch value of $\al$ turns out to coincide with
the MOND scale \cite{mond,McGaugh} is intriguing. To start thinking about
possible connections, it pays to recall that $\al$ is an estimate of the
relative deceleration of wall regions, at finite infinity where space
is still expanding, with respect to the cosmological volume average at
any epoch. It is certainly an acceleration scale at which dynamical
gradients in the Ricci curvature of space are likely to affect dynamics of
particles between walls and void regions.

Since we are no longer dealing with asymptotically flat geometries
with exact timelike Killing vectors, it is quite possible that the solution
to the Kepler problem for bound geodesics in galaxies should be modified.
Therefore the possibility that the MOND phenomenon is related is a
reasonable hypothesis. However,
the estimate we have made of the relative deceleration scale, $\al$,
refers specifically to the finite infinity scale relative to
the volume average. Given that the outskirts of galaxies are expected
to lie within finite infinity, it means that no detailed quantitative
comparisons can be made until the transition zone around finite infinity
is more directly modelled.

I will not attempt to look at the problem of the rotation curves of
galaxies in the present paper, but will make a few observations
that would follow if the effects of MOND are simply a modification of
Newtonian dynamics in a static background which arise from dynamical
density gradients affecting the Ricci curvature of space.

The first important point is that, since we are dealing with
an effect which is most pronounced between walls and voids, then the
distance of a galaxy to a finite infinity boundary should play
a role, at least na\"{\i}vely. One might expect more pronounced effects
for galaxies in filaments \cite{foot13} as compared to rich clusters
of galaxies in thick walls, where the distance to finite infinity is greater.
In fact, rich clusters of galaxies often tend to harbour elliptical galaxies
for which the MOND results differ little from Newtonian expectations. But
apart from this, it appears that for the nearby redshifts over which it is
tested, the MOND scale \cite{mond} is uniform for many different galaxy
environments. This runs counter to na\"ive intuition about the distance
to finite infinity, and suggests
that some key physical insight remains to be found.

The second point is that, although the magnitude of the relative deceleration
is larger at higher redshifts, the frequency of voids is also less. There
may be a trade--off between these two competing factors as to an optimal
redshift at which dynamical effects would be most evident. For the best--fit
parameters \cite{LNW}, the void fraction reaches 50\% at a redshift $z=1.52$
when $\al\simeq3.4\times10^{-10}$ms$^{-2}$. It is possible that effects
resulting from the relative deceleration of the background may be important
for structure formation. Toy model calculations
based on exact inhomogeneous solutions of Einstein's equations show that
the nonlinear treatment of inhomogeneities in general relativity can
considerably enhance the rate of structure formation \cite{Bolejko}.

Given that the expected distance between finite infinity and the outskirts
of a galaxy does not suggest a direct link between $\alpha_0$ and
$\alpha_{mond}$, the close match between $\alpha_0$ and $\alpha_{mond}$
may be purely coincidental, even if both effects are related to Ricci
curvature gradients which are typically of the same order of magnitude.
As no direct link to MOND has yet been established, it is not
possible to say whether the redshift dependence of $\alpha$ should be
linked to a redshift dependence in the MOND scale. MOND is
purely empirical and the fitting of rotation curves requires both an
acceleration scale and an empirical function, $\mu(x)$, which interpolates
between the Newtonian and ``modified dynamics'' regimes. Any first
principles treatment would need to describe the transition zone between
finite infinity and its interior, which might conceivably be related
empirically to the interpolating function.
We are at an early stage of gathering the pieces of a puzzle, where
all observations have to be treated with rigorous scrutiny, and where
no firm conclusions should be drawn until a compelling theoretical
case is assembled.

Finally, it should be noted that the best--fit parameters
of Ref.\ \cite{LNW} indicate that nonbaryonic dark matter is likely
to be the dominant component of matter in the universe by mass, even if
the relative fraction is generally somewhat lower than in the $\Lambda$CDM
model. If there is any link to MOND, then it is the phenomenology of
MOND that one might hope to explain, rather than an alternative to
nonbaryonic dark matter.

\section{Weyl curvature and initial conditions\label{Weyl}}

The cosmological equivalence principle has been applied here principally
on the scale of macroscopic cosmological averages. A question remains as
to what extent we should take it as a universal principle? Given that
the universe is well approximated by frames of the sort (\ref{cif})
at the earliest epochs which are observationally tested, such as the epochs
relevant to big bang nucleosynthesis, it is certainly
tempting to try to apply the CEP on arbitrarily small scales within
the past light cone at the earliest epochs. For any perfect fluid the
pressure is determined by the density and no modification of the CEP is
required. However, if one is dealing with spinning fluids or gravitational
waves which cannot be treated by a single collective scalar degree of
freedom, then modifications to
the stated principle would be necessary. Is the process of
``getting rid of local Ricci curvature'' to recalibrate rods and clocks
so that the effects of gravity disappear in a small region,
as the equivalence principle demands, related to the mathematical problem
of Ricci flow? This is certainly a possibility, which
has been discussed in the context of cosmological averaging by Buchert
and Carfora \cite{BC2}.

As far as the rest of the curvature degrees of freedom are concerned, the
CEP in fact incorporates a strong statement about average Weyl curvature,
since the CIF (\ref{cif}) has vanishing Weyl curvature. Of course, Weyl
curvature does not vanish everywhere identically -- it is nonzero in the
vicinity of collapsed structures such as stars and black holes, or in
gravitational wave ripples. However,
the statement of the CEP means that average effects of Weyl curvature need
not be considered in the calibration of rods and clocks of generic
cosmological frames. This ``average Weyl curvature condition'' may at
first sight seem stronger than Penrose's ``Weyl curvature hypothesis''
\cite{Penrose} that the universe {\em began} in an initial state with
total Weyl curvature exactly zero. However, at the epoch of last scattering
the universe was very close to a standard FLRW model, and all FLRW models
have vanishing Weyl curvature. Thus the ``average Weyl curvature condition''
incorporated within the CEP could be viewed as a relatively weak
phenomenological statement about the evolution of a universe whose initial
state at last scattering is consistent with the predictions of inflation.
Alternatively, it is also consistent with any other cosmological scenario
which solves the flatness and horizon problems to give a close to spatially
flat FLRW universe with near scale--free perturbations by the epoch of big
bang nucleosynthesis, whether such a scenario satisfies Penrose's Weyl
curvature hypothesis or not.

It is my view, however, that in searching for potential scenarios for
initial conditions, Penrose's Weyl curvature hypothesis needs to
be taken seriously, and could be related to a generalised cosmological
equivalence principle, if it can be formulated to apply at earlier epochs
in the very early universe. To understand my rationale let us recall that
the Weyl curvature tensor includes any nonlocal curvature in a manifold,
whereas the Ricci tensor encodes purely local curvature since it is
directly related to the energy--momentum tensor via the Einstein equations.

Since general relativity is a local causal theory, given that the
observable universe was in a global state very close to a FLRW geometry
with zero Weyl curvature at last scattering, then the only Weyl curvature
we are allowed today is that which has accumulated by local
causal processes within the past light cone at any event: in particular, by
gravitational collapse and production of gravitational waves. The Weyl
tensor encodes tidal curvature information
on local scales which grows as matter clumps. On the large scales where
the universe is still expanding, Weyl curvature cannot be important
in defining the average geometry, since the universe
has only had a finite time over which to evolve from its state
at last scattering. This is also the reason why there is a statistical
scale of average homogeneity.

If a version of the CEP can be taken
to apply at even the earliest epochs, then it amounts to the
statement that even throughout its earliest history the background universe
contains no ``nonlocal curvature'' that cannot have evolved causally within
the past light cone at any event. For as far back as a 4--dimensional
spacetime continuum has any meaning it would then make sense to be able to
choose a CIF in which the average effects of density are volume--contracting.
Large classes of models, such as Bianchi models with anisotropic flows, would
be cosmologically irrelevant. If such a CEP should survive
through the epoch of inflation or any other very early universe scenario, then
it would coincide with Penrose's Weyl curvature hypothesis. Ultimately these
conceptual issues might inform quantum gravity
and quantum cosmology.

\section{Discussion\label{conclude}}

In this paper I have extended the strong equivalence principle to
account for the average effect of the density of matter in the definition
and relative calibration of clocks in inertial frames on cosmological
scales. Since the resulting cosmological
equivalence principle relates the single scalar degree of freedom of Newtonian
gravity to the framework of general relativity, it may provide a means to
better understand the calibration of cosmological weak
fields once density perturbations have grown large to form a universe
that is very inhomogeneous on scales of tens of megaparsecs. It
should thereby give a setting for better understanding the Newtonian limit
in the dynamical situation of cosmology. The numerical estimate of
the relative deceleration between observers in the walls around galaxy
clusters and volume-average
observers in voids, typically of order $10^{-10}$ms$^{-2}$, is acceptably small
for weak field scales and yet leads cumulatively to the present epoch clock
rate variance of 38\% found in Refs.\ \cite{clocks,LNW}. Intriguingly, at
redshifts $z\lsim0.25$ the relative deceleration required by the CEP
coincides precisely with the empirical acceleration scale of MOND
\cite{mond,McGaugh}.

At a conceptual level I have attempted to present a framework for the
consistent definition of average inertial frames in relation to average
dynamically--varying matter densities in cosmological general relativity.
The hope is that the cosmological equivalence principle is thereby a
key step to the incorporation of Mach's principle into general relativity,
in the way that Einstein intended but never quite realised.
Mach's principle is most commonly invoked in distinguishing inertial frames
from rotating frames \cite{BKL,BKL2,Schmid,wave}. What is studied
here is a different aspect of Mach's principle: the role of the
average volume deceleration of the local geometry in defining the
standard of time of inertial frames. In the absence of a timelike Killing
vector, the evolution of the average density provides a relevant clock.
To fully incorporate Mach's principle in general relativity,
it is of course necessary to deal with the other dynamical gravitational
degrees of freedom which can affect the distinction of inertial frames
from rotating ones, such as gravitational waves \cite{wave}.

In this paper I have expounded the
view that to deal with the volume--contracting average dynamical effects of
matter density, a reduction to a frame (\ref{cif}) is the relevant step
in the normalisation of gravitational energy before the final step to a
static Minkowski space. It is quite possible that when average degrees of
freedom in addition to the scalar Ricci curvature are considered, in order
to deal with gravitational waves and spinning matter fluids, there
are other steps in the relevant relative calibrations of rods and clocks.
After all, energy, momenta, and angular momenta are only defined with
respect to a frame. In the dynamical regime of general relativity the
question arises as to which collective average frames have physical
utility in the absence of exact symmetries described by Killing
vectors. My view is that a truly deep understanding of quasilocal
gravitational energy and momentum is still to be found, but the path to
such enlightenment requires a better conceptual understanding of the
equivalence principle in application to collective dynamical degrees
of freedom of matter fields.

Historically speaking, in the early stages of the development of general
relativity Einstein did not fully appreciate the dynamical importance of the
energy and momentum of spacetime itself. Spacetime is inevitably dynamical
for matter obeying the strong energy condition. Einstein's first journey
through the conceptual landscape of cosmological general
relativity had him worrying about boundary conditions at spatial
infinity \cite{esu}, as he overlooked the possibility that the
universe had a beginning.

Since general relativity is causal the geometry at any event can only
depend on events within its past light cone, and is independent of what lies
beyond the particle horizon. Thus boundary conditions at spatial infinity
beyond the particle horizon are physically irrelevant if the universe had a
beginning, a possibility that Einstein did not consider when he first
formulated his static universe \cite{esu}. For a universe like ours which
had a beginning, the initial conditions are of vital importance in
determining the relevant weighted average of the apparent motions,
as I have discussed in Sec.\ \ref{Weyl}.

The conceptual journey discussed in this paper arose in an effort to model
the universe more realistically \cite{clocks,sol,LNW}, to account for the
structure we actually observe, by realising that the quasilocal gravitational
energy of a dynamical spacetime geometry -- which has real effects on the
calibration of clocks -- should be an essential feature of a universe
with large dynamical density gradients. If successful, this will eliminate
the need for a cosmological constant or other fluidlike vacuum energy as
the source of ``dark energy'', but it still leaves the other cosmological
problem, why $\Lambda=0$, unsolved. If we take the strong equivalence
principle literally then $\Lambda$ must be zero since otherwise
we could not have a vacuum Minkowski spacetime for our local inertial frames.
My own personal view is that quantum field theoretic calculations based in a
flat spacetime which suggest that $\Lambda\goesas M\ns{Planck}^4$
miss the mark, because the spacetime vacuum cannot be understood
without accounting for the intrinsically dynamic nature of spacetime.
It is not a problem for flat space quantum field theory. While the
cosmological constant problem is no doubt a problem for quantum
gravity, I believe that quantum gravity research might benefit
from a more physical understanding of dynamical gravitational energy
and the equivalence principle.

\medskip {\bf Acknowledgement} I wish to thank many people for
discussions and correspondence including in particular Thomas Buchert, Syksy
R\"as\"anen, Frederic Hessman, Herbert Balasin, Helmut Rumpf, Peter Aichelburg,
Lars Andersson and Stacy McGaugh. I warmly
thank Prof.\ Remo Ruffini and ICRANet for support and hospitality while
the paper was completed.


\begin{thebibliography}{66}
\bibitem{foot1}
{Alternatively, the special
relativistic formula is exact if applied to the radial 3-velocity
extracted from the vector field obtained by parallel propagation of the
emitter 4--velocity to the source along the relevant null geodesic.
It has components $U_{\parallel}^{\hat a}=c\left[(a\Z0^2+a^2)/(2a\Z0a),
(a\Z0^2-a^2)/(2a\Z0a),0,0\right]$ in an orthonormal frame at the source.}
\bibitem{foot2}
{Operationally this is possible by
cutting a volume out of a time slice of a generic Milne congruence
\cite{Milne}, depending on some desired accuracy.}
\bibitem{foot3}
{The details of the
synchronisation of clocks and the tethering will not be considered in
detail as I will use this as a conceptual, rather than a calculational
device. However, in the analogous ``tethered galaxy problem''
\cite{Harrison} a number of aspects of the solution are best understood
using radar time \cite{Clavering}.}
\bibitem{foot5}
{Some progress has been made in understanding the origin
of spatial curvature in Zalaletdinov's scheme \cite{CPZ}, and also in
its relation to Buchert's scheme \cite{PS}, and its relationship to
perturbation theory at early times \cite{Paranjape}. The scheme has
also been applied to bound systems \cite{vandenHoogen}. However, with regard
to the specific question of vorticity in the fully non-linear regime, work
remains to be done.}
\bibitem{foot6}
{Since CIFs like LIFs are 4-dimensional frames this
operational approach demands that we consider near--to-spacelike slices
which have some small thickness.}
\bibitem{foot7}
{Similarly to the results of Refs.\ \cite{clocks,sol,LNW}, in LTB
models the volume average deceleration can be positive, despite there being
apparent acceleration in the radial Hubble flow relative to the central
observer \cite{BA}.}
\bibitem{foot9}
{In the notation of Buchert and Carfora \cite{BC3}, $\QQ\Z{\cal M}=0$
or $\de^2H\Z{\cal M}=\frn13\ave{\si^2}\Z{\cal M}$, $\QQ\Z{\cal E}
=0$ or $\de^2H\Z{\cal E}=\frn13\ave{\si^2}\Z{\cal E}$, and
$\QQ\Z{\cal D}=6\la\Z{\cal M}(1-\la\Z{\cal M})(H\Z{\cal E}-H\Z{\cal M})^2$.
Relative to the notation of \cite{clocks,sol},
$H\Z{\cal M}= H\ns{w}$, $H\Z{\cal E}= H\ns{v}$ and
$\la\Z{\cal M}=1-\fv$. However, the limit $H\Z{\cal M}\to0$ considered
in Ref.\ \cite{BC3} is not considered in my work, as comparing expansion of
a finite infinity region relative to the volume average is the
important feature in determining relative clock rates.}
\bibitem{foot10}
{There is a small inconsequential error in the last term of
eq.\ (57) of Ref.\ \cite{clocks}. This should be corrected to read
$H=\dots=\bH\left[2\gb-\frn32-{\gb\Omb[(4+\Omb)\gb^2-8\gb+3]\over2(1-\gb^2
\Omb)}\right]$. Furthermore, in eq.\ (4) of Ref.\ \cite{clocks},
$\ave{\si}^2$ should read $\ave{\si^2}$.}
\bibitem{foot12}
{It is sometimes remarked that the MOND scale and Pioneer anomaly
are ``of the same order'' \cite{mond}, ignoring a factor of 5--8 if $H\Z0=75
\kmsMpc$. Our value of the Hubble constant exacerbates the difference, to
make the MOND scale one order of magnitude smaller than the Pioneer anomaly
scale.}
\bibitem{foot13}
{Since the Hubble flow starts just
1.5--2Mpc from the Local Group it appears that the Local Group and nearby
dwarf galaxies actually belong to a filament.}
\end{thebibliography}
\end{document}